\documentclass[aps,prx,twocolumn,amsmath,nofootinbib,amssymb,superscriptaddress,10pt]{revtex4-1}
\usepackage{graphicx}
\usepackage{ulem}
\usepackage{soul}
\usepackage[colorlinks, linkcolor=blue]{hyperref}
\usepackage{verbatim}
\usepackage{esint}
\renewcommand{\vec}[1]{\boldsymbol{#1}}
\def \k {{\vec k}}

\def \ve {\varepsilon}
\def \r {{\vec r}}
\def \A {\mathrm{A}}
\def \B {\mathrm{B}}

\def \el {\ell}

\def \q {{\vec q}}

\def \l {\ell}
\def \ve {\varepsilon}

\def \beq {\begin{eqnarray}}
\def \eeq {\end{eqnarray}}
\def \tn {\textnormal}

\def \ua {\uparrow}
\def \da {\downarrow}
\begin{document}

\title{Intrinsic superconducting instabilities of a solvable model for an incoherent metal}

\author{Debanjan Chowdhury}
\affiliation{Department of Physics, Cornell University, Ithaca, New York 14853, USA.}
\affiliation{Department of Physics, Massachusetts Institute of Technology, Cambridge, Massachusetts
02139, USA.}
\author{Erez Berg}
\affiliation{Department of Condensed Matter Physics, Weizmann Institute of Science, Rehovot, 76100, Israel.}
\begin{abstract}
We construct a family of translationally invariant lattice models with a large number ($N$) of orbitals at every site coupled together via single electron tunneling. By tuning the relative strength of the electronic bandwidth and on-site interactions, that have a modified Sachdev-Ye-Kitaev (SYK) form, we demonstrate a number of unusual features at strong coupling and in the large$-N$ limit. We find examples of (i) an intrinsic non-BCS superconducting instability arising out of an incoherent non-Fermi liquid metal, and, (ii) an instability of an incipient heavy Fermi liquid metal to superconductivity with transition temperatures comparable to its renormalized bandwidth. At strong-coupling, these solvable models display pairing instabilities that are not driven by any special ``nesting" properties associated with an underlying Fermi surface.
\end{abstract}
\maketitle
\section{Introduction}
\label{intro}

The search for high-temperature superconductivity in complex electronic materials continues to be at the heart of modern condensed matter physics. One of the most successful and well understood theories of a non-trivial collective many-body effect is the Bardeen-Cooper-Schrieffer (BCS) theory of phonon mediated superconductivity in conventional metals such as aluminum. However, most of the high-temperature superconducting materials evolve out of a metallic state that is highly unconventional and can not be described within Fermi-liquid (FL) theory with long-lived quasiparticles. Some of the most notable examples of such materials are the copper-oxide based (``cuprate") \cite{Keimer15}, iron-pnictide (chalcogenide) based \cite{scalapino} and certain rare-earth element based \cite{Stewart} compounds, where the parent state of the superconductor is a non-Fermi liquid (NFL) metal. At the same time, it is also likely the case that superconducting pairing in these materials occurs through a purely electronic mechanism. 

While not all of the different families of NFL metals display an identical phenomenology, they share a number of peculiar features. These include short single-particle lifetimes \cite{zxs,johnson}, a broad regime of anomalous power-law transport seemingly at odds with expectations in a Fermi liquid \cite{Takagi,Marel,Taillefer1,hussey} and absence of any characteristic crossovers through the ``Mott-Ioffe-Regal" (MIR) limit where the electronic mean-free path becomes of the order of the lattice spacing \cite{Hussey04}. These systems are often called ``bad-metals" \cite{SAK95}, or ``strange-metals" and the nature of their pairing instabilities, driven purely as a result of repulsive electronic interactions remains poorly understood. Finding concrete examples of models where the emergence of NFL behavior and superconductivity can be analyzed through reliable theoretical means is thus of paramount importance.

Over the past few decades, a few different frameworks have been studied to describe the properties of NFL metals. Focusing specifically on translationally invariant models with local electronic interactions, a recent approach that has been successful in capturing some of the NFL phenomenology over a broad range of temperature and energy scales relies on using a ``solvable" (but artificial) building block, the Sachdev-Ye-Kitaev (SYK) model \cite{SY,kitaev_talk,Maldacena2016,kitaevsuh,Parcollet1,Parcollet2}. The SYK model is a $(0+1)-$dimensional model that consists of a large number of orbitals interacting with random all-to-all interactions on a single site. The transport properties of a higher dimensional lattice generalization of such SYK islands with strong disorder have been studied in a number of different settings \cite{Gu17,SS17,Balents,Zhang17,SSmagneto,DVK17,hongyao}. 

In a previous paper \cite{DCsyk}, we constructed a family of models with {\it exact} translational symmetry and on-site SYK-like interactions. This allowed us to study the fate of electronic quasiparticles and sharply defined Fermi surfaces, in the regime of strong interactions. The aim of the present paper is to further extend these models to study the possible onset of pairing, mediated by the same interactions that are also responsible for destroying the quasiparticles and the underlying Fermi surface (i.e. we do {\it not} include any ``bare" non-SYK attractive/repulsive interactions). We will explicitly construct models where attraction is effectively generated in the pairing channel (and possibly other channels),  and are unlike the conventional weak-coupling BCS type instabilities.  

In this paper, we modify the one-band translationally invariant model considered by us in Ref.~\cite{DCsyk}, to include an additional spin label and local $SU(2)$ invariant interactions. If the bandwidth is given by $W$ and the typical interaction strength is $J$, at the level of the large-$N$ saddle-point equations, we find that the system crosses over at a temperature $T_{\tn{coh}}\sim W^2/J$ from a low-temperature Landau Fermi liquid ground state to locally quantum critical non-Fermi liquid state, where the Fermi surface is completely destroyed. The dc resistivity crosses over from $\rho \sim T^2$ at $T\ll T_{\tn{coh}}$ to $\rho \sim T$ at $T\gg T_{\tn{coh}}$; the value of the resistivity at the crossover scale $(T\sim T_{\tn{coh}})$ is $\rho\approx h/Ne^2$. Depending on the fine details of the model related to the nature of correlations between the on-site interaction matrix elements, we can obtain two qualitatively distinct outcomes. For one of the cases, we find that the high-temperature NFL metal becomes unstable to superconductivity with an on-site, spin-singlet order parameter. The transition temperature is set by the large on-site interaction scale ($T_c\sim J$); long-range order sets in as a result of the Josephson coupling (${\cal{J}}\sim NT_{\tn{coh}}$) between nearest neighbor sites. For the other case, we find that the high-temperature incoherent metal is stable against pairing but the incoherent excitations in this regime give rise to a significant enhancement in the strength of pairing correlations as a function of decreasing temperature. Moreover, across the crossover scale $T_{\tn{coh}}$ to the incipient Fermi liquid regime, there is a pairing instability with a transition temperature that is set by the same scale, $T_c\sim T_{\tn{coh}}$ (which is also the renormalized bandwidth). In neither of the two cases does the pairing instability arise as a result of the conventional ``Cooper-logarithm". 

The remainder of this paper is organized as follows. In Section \ref{prelim}, we introduce the basic setup of the problem. Section \ref{model} contains a discussion of our model of electrons with $N$ orbitals at every site, where each orbital has an additional spin$-1/2$ label, and obtain the large$-N$ saddle-point equations; the results of this subsection are qualitatively similar to Ref.~\cite{DCsyk}. In Section \ref{instab}, we analyze the Bethe-Salpeter equations in the pairing channel at the same leading order in $1/N$ for the models introduced in Sec.~\ref{model}. Sections \ref{MA} and \ref{MB} contain a detailed discussion of the pairing instabilities for the two distinct families of models. We conclude with a discussion and a future outlook in Section \ref{disc}. In Appendix \ref{sykq}, we study the same problem using SYK$_q$ as a building block for $q>4$ to highlight the interesting underlying structure of the solutions.

\section{Preliminaries}
\label{prelim}

\subsection{Model} 
\label{model}

Our starting point will be a  generalization of the model introduced in Ref.~\cite{DCsyk}, written in terms of complex fermions with an orbital ($i=1,...,N$) and spin ($s=\uparrow,\downarrow$) labels. The Hamiltonian with purely on-site interactions, that preserves a global $U(1)$ charge-conservation and a global $SU(2)$ spin-symmetry{\footnote{$H_J$ is not the most general form of the Hamiltonian with $SU(2)$ symmetry but is chosen to have the specific form for simplicity.}} is given by ($\r$ is defined on a $d-$dimensional lattice),
\begin{subequations}
\beq
H &=& H^{}_t + H^{}_J + H^{}_P ~,\label{ham}\\  
H^{}_t &=& \sum_{\r,\r'}\sum_{i,\{s_1=\ua,\da\}}(-t_{\r\r'} - \mu \delta_{\r\r'})~ c^\dagger_{\r i s_1} c_{\r' i s_1},\\
H^{}_J &=& \frac{1}{4N^{3/2}}\sum_\r \sum_{i,j,k,\el} \sum_{\{s_i=\ua,\da\}}J_{ijkl}~ c^\dagger_{\r is_1} c^\dagger_{\r js_2} c_{\r ks_2} c_{\el \r s_1}, \label{hu}\nonumber\\ \\
H^{}_P &=& \frac{1}{N}\sum_\r \sum_{i,j} U c^\dagger_{\r is} c^\dagger_{\r i-s} c_{\r j-s} c_{\r js}.\label{hp}
\eeq
\end{subequations}
The hopping strengths, $t_{\r\r'}$, in $H_t$ are assumed to be identical for each orbital and $\mu$ represents the chemical potential. We have introduced an additional spin label in order to allow us to distinguish between spin-singlet vs. triplet pairing instabilities. 
We assume that the interaction strengths $J_{ijkl}$ are drawn from an independent random distribution with $\overline{J_{ijkl}}=0$ and $\overline{J_{ijkl}^2}=J^2$ and require that $J_{ijkl}$ be antisymmetric with respect to changing the following indices: $J_{ijkl} = -J_{jikl} = -J_{ijlk}$. In addition, choosing these strengths to be real leads to the condition $J_{ijkl} = J_{klij}$. Note that just as in Ref.~\cite{DCsyk}, we are constructing a translationally invariant model where the interaction matrix elements, $J_{ijkl}$, are identical at every site (i.e. independent of the site label, $\r$) and the momentum $\k$ is thus a good quantum number. This aspect of our model requires care when carrying out disorder-averaging, as was emphasized in Ref.~\cite{DCsyk}. Finally, we have also included an on-site `pair-hopping' term, $H^{}_P$, with uniform $U>0$, which suppresses the tendency towards on-site pairing.  

Let us now introduce additional structure on the precise form of the interaction matrix elements ($J_{ijk\el}$) by considering two {\it distinct} scenarios. Consider the permutation symmetries under exchanging the second and third index for `Model-$\A$' (`Model-$\B$') to be of the form, $J_{ijkl} = J_{ikjl}$ ($J_{ijkl} = -J_{ikjl}$). This subtle distinction between the two models leads to significant differences for the resulting instabilities. It is worth noting that Model-$\A$ can only be defined for the version of the SYK model written for complex, but not Majorana, fermions. Interestingly, including contributions from $H^{}_P$ by making $U$ large for Model-$\A$ allows us to access the physics described by Model-$\B$, as we shall demonstrate below.  

\begin{figure}[h]
\begin{center}
\includegraphics[scale=0.3]{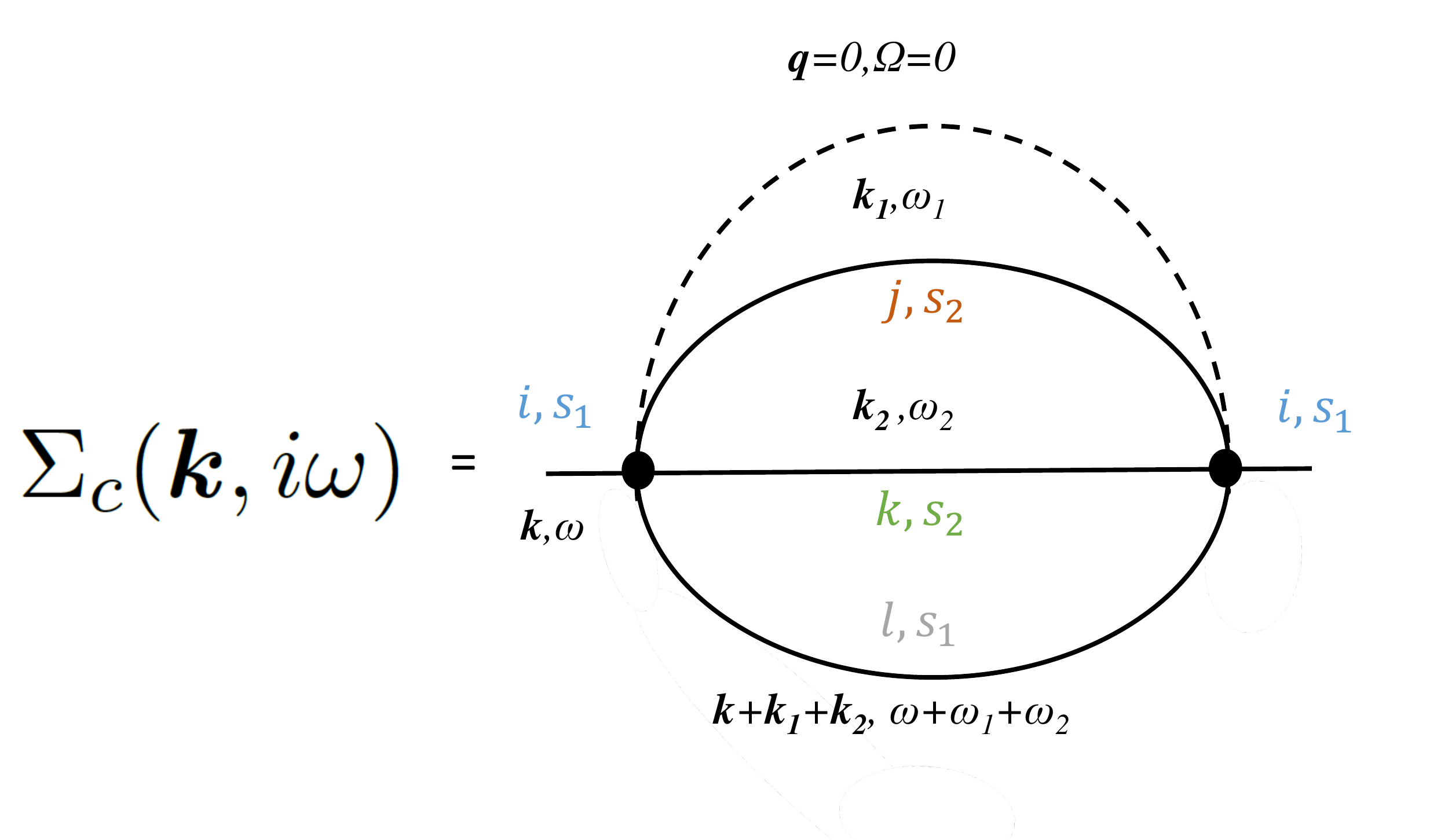}
\end{center}
\caption{The electronic self-energy for orbital $i$ with spin-label $s_1$. The solid black lines represent fully dressed Green’s functions, $G(\k,i\omega)$ and the dashed lines represent the $J^2$ contraction.}
\label{se}
\end{figure}

We are interested in the large$-N$ saddle point solution for the model described in Eqn.~\ref{ham}. These equations turn out to be identical for both Model-$\A$ and $\B$ and can be expressed in terms of the usual self-consistent set of equations for the electron Green's function (i.e. `watermelon' diagrams; see Fig.~\ref{se}) \cite{DCsyk},
\beq
G(\k,i\omega) &=& \frac{1}{i\omega - \ve_\k - \Sigma(\k,i\omega)}, \label{saddle}\\
\Sigma(\k,i\omega) &=& -J^2\int_{\k_1}\int_{\omega_1} G(\k_1,i\omega_1)~\Pi(\k+\k_1,i\omega+i\omega_1),\nonumber\\
\Pi(\q,i\Omega) &=& \int_{\k}\int_{\omega} G(\k,i\omega)~G(\k+\q,i\omega+i\Omega).
\eeq
Note that $H^{}_P$ does not enter the above equations at this order in large$-N$. The solution for the Green's function in the strong-coupling limit ($J\gg W$) is given by,
\begin{equation}
G(\k,i\omega) \sim 
\begin{cases}
\frac{Z}{i\omega - Z \overline\varepsilon_\k + i \alpha \nu_0^2 J|\omega|^2\ln(\frac{W^*}{|\omega|})\mathrm{sgn}(\omega)}, ~\omega \ll W^*,\\
\frac{i\mathrm{sgn}(\omega)}{\sqrt{J |\omega|}} - B(\omega)\frac{\varepsilon_{\k}}{J |\omega|}, ~~~~~W^* \ll \omega \ll J,
\end{cases}
\label{limits}
\end{equation}
where $W^*\sim W^2/J$ is the renormalized bandwidth, $Z \sim 1/(\nu_0 J)$ is the quasiparticle residue ($\nu_0\sim1/W$ is the single-particle density of states, where we use units where the lattice spacing $a=1$), $\overline\varepsilon_\k$ is the renormalized dispersion 
($\overline\varepsilon_\k / \varepsilon_\k$ is of order unity in the strong coupling limit), 
and $\alpha$ is a number of order unity (the $\log$ appears only in two-dimensions). The factor of $B(\omega)$, that descends from the ``spectral asymmetry", is a constant independent of frequency but whose value depends only on the sign of $\omega$. At strong coupling, we thus observe a crossover from a high temperature (or, energy) incoherent metal (IM) without any momentum-space structure to a low temperature Fermi liquid (FL) at a characteristic scale of $T_{\tn{coh}}\sim W^*$. On the other hand, at weak-coupling ($J\ll W$), the system remains a FL at all temperatures with $Z\sim 1 - (\nu_0J)^2$. However, the existence of the above self-consistent solution does not preclude the possibility of a finite temperature instability of the metallic states. This will be the topic of our study in the next few sections. 

\subsection{Bethe-Salpeter Equations}
\label{instab}

Let us now investigate the possible instabilities in the particle-particle ($pp$) channel for the metallic states obtained within the large$-N$ analysis in the previous section. We shall study and contrast the properties of the model in Eqn.~\ref{ham} in the following limits:

\begin{itemize}
    \item Model-$\A$ (i.e. $J_{ijkl} = J_{ikjl}$),
    \item Model-$\B$ (i.e. $J_{ijkl} = -J_{ikjl}$).
\end{itemize}

To begin the discussion, we note that the matrix-elements, $J_{ijkl}$, have both attractive and repulsive components and no net attraction on average (i.e. $\overline{J_{ijkl}}=0$). We may thus be tempted to conclude that the model as defined in Eqn.~\ref{ham}-\ref{hp} has no interaction-driven instability in any channel. However, this naive expectation is {\it incorrect}. Instead, as we shall explicitly demonstrate below, attraction can be generated at the same leading order in $N$ but at higher order in $J^2$.  

\begin{figure}[h]
\begin{center}
\includegraphics[scale=0.17]{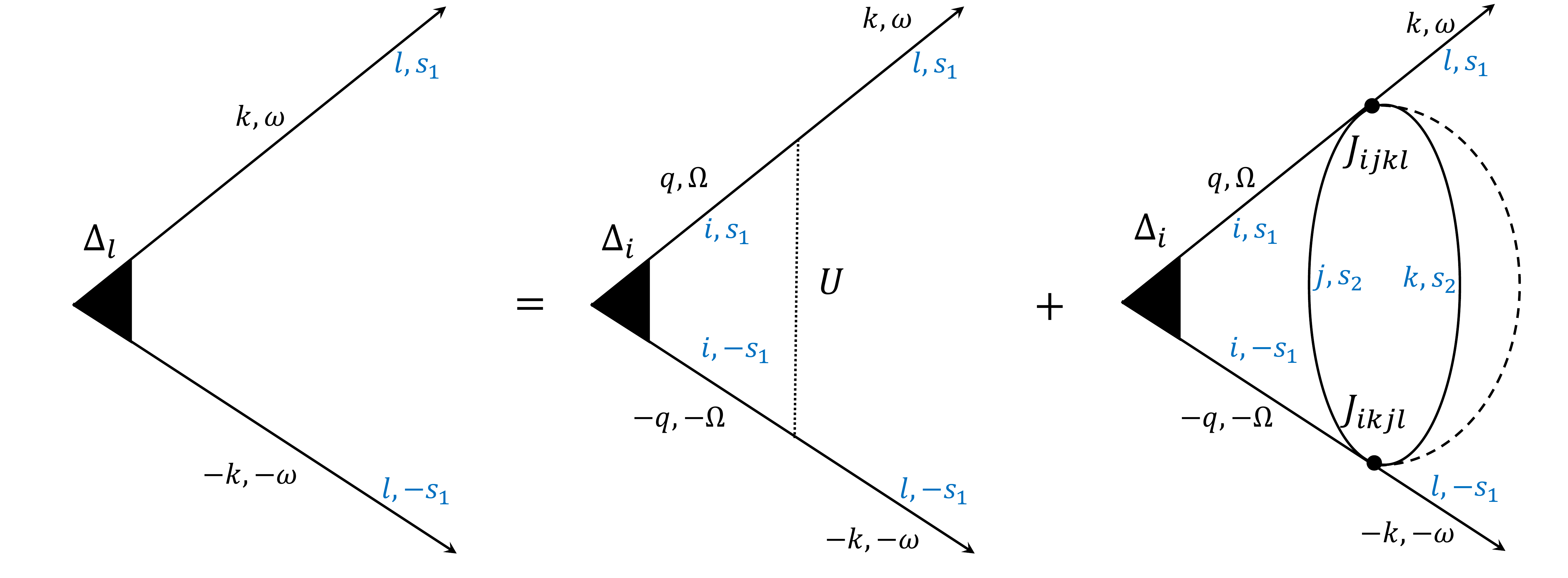}
\end{center}
\caption{The Bethe-Salpeter equation for the pairing vertex in the particle-particle channel. Solid lines denote the fully dressed electron propagators, the dotted line denotes the $U-$interaction vertex and the dashed line denotes the disorder contraction $\overline{J_{ijkl}J_{ikjl}}$.} 
\label{pv}
\end{figure}

Consider the following vertex in the (spin-singlet) $pp$ channel,
\beq
\Delta_{ij}(\r-\r') &\equiv& \langle \epsilon_{s_1s_2}~ c_{\r is_1} c_{\r' js_2}\rangle,
\eeq
and let us study the linearized Bethe-Salpeter equations for the above vertices by going to $O(J^2)$ (see Fig.~\ref{pv}). From a simple counting argument, it is immediately clear that the intra-orbital component of $\Delta_{ij}\propto\delta_{ij}$ is not suppressed in $1/N$ (while the inter-orbital component will be suppressed). Focusing specifically on model-$\A$ in the presence of a finite $U$, we find that at the same leading order in $N$, the bare pair-hopping term suppresses on-site pairing. 

The Bethe-Salpeter equations for model$-\A$ and model$-\B$ at zero external center-of-mass momentum and in the spin-singlet channel are then of the form,
\begin{widetext}
\begin{subequations}
\beq
\tn{Model}-\A: &&\Delta_\l(\k,\omega) = - T\sum_{\Omega}\int_\q\Delta_i(\q,\Omega)~G_i(\q,i\Omega)~G_i(-\q,-i\Omega)\bigg[U + J^2~\Pi(\k-\q,i\omega-i\Omega)\bigg],\label{BS1a}\\
\tn{Model}-\B: &&\Delta_\l(\k,\omega) = -  T\sum_{\Omega}\int_\q\Delta_i(\q,\Omega)~G_i(\q,i\Omega)~G_i(-\q,-i\Omega)\bigg[U - J^2~\Pi(\k-\q,i\omega-i\Omega)\bigg].\label{BS1b}
\label{BS2a}
\eeq
\end{subequations}
\end{widetext}

Before analyzing the above equations in detail, we note in passing that we could have also assumed the interaction matrix elements to be uncorrelated in the following sense: $\overline{J_{ijkl}J_{ikjl}} = 0$. The term proportional to $J^2~\Pi$ would then be absent{\footnote{The ladder insertion in the Bethe-Salpeter equation in the pairing channel involves the contraction $\overline{J_{ijkl}J_{ikjl}}$. On the other hand, in the particle-hole channel (to be discussed later), the contraction would be of the form $\overline{J_{ijkl}J_{ijkl}}$. }} from the above equations and the model in Eqn.~\ref{ham}-\ref{hp} would not have any on-site pairing instability at the leading order in $1/N$. It is still possible to study the onset of pairing by introducing an explicit infinitesimal attractive interaction \cite{Patel}. 

\section{Model$-\A$: Single-site instability}
\label{MA}

As discussed in Sec.~\ref{model}, the incoherent metal for $T\gg T_{\tn{coh}}$ has SYK-like correlations with a completely local electron Green's function and corrections that are perturbative in the strength of hopping. Solutions to the Bethe-Salpeter equations, if any, have to thus emerge at the level of the single site and there is no `Fermi surface' driven instability.  Therefore, without any loss of generality, we ignore the momentum dependence of the Green's functions. Moreover, in this regime let us set $U=0$ and begin by asking if a finite $J$ can give rise to an instability. The equations for model$-\A$ then simplify to,
\beq
\Delta_\l(\omega) &=& -J^2 T\sum_{\Omega} \Delta_i(\Omega)~G_i(i\Omega)~G_i(-i\Omega)~\Pi(i\omega-i\Omega),\label{gap1}\nonumber\\ 
\eeq
where $G_i(i\Omega) = 1/\left[i\Omega - \Sigma(i\Omega)\right]$, and $\Sigma(i\Omega)$ is given by the solution of Eqs.~(\ref{saddle}). Since the only energy scale in the problem is set by $J$, it is useful to scale both $\Omega$, $T$, and $\Sigma$ by $J$, obtaining
\beq
\Delta_\l(\widetilde{\omega}) &=& -\widetilde{T}\sum_{\widetilde{\Omega}} \Delta_i(\widetilde{\Omega})~\widetilde{G}_i(i\widetilde{\Omega})~\widetilde{G}_i(-i\widetilde{\Omega})~\widetilde{\Pi}(i\widetilde{\omega}-i\widetilde{\Omega}).\label{gap2}\nonumber\\ 
\eeq
Here, tildes represent dimensionless quantities obtained by rescaling by $J$, e.g., $\widetilde{\Omega} = \Omega/J = \pi(2n+1)\widetilde{T}$ where $\widetilde{T} = T/J$ and $n\in \mathbb{Z}$, $\widetilde{G}_i(\Omega) = 1/\left[i\widetilde{\Omega}-\widetilde{\Sigma}(\widetilde{\Omega})\right]$, etc., and $\widetilde{\Pi}(\widetilde{\Omega}) = \widetilde{T}\sum_{\widetilde{\omega}} \widetilde{G}_i(i\widetilde\omega) \widetilde{G}_i(i\widetilde{\Omega}+i\widetilde{\omega})$. Eqn.~(\ref{gap2}) thus has a solution when the linear operator $M(\Delta)$  defined by the right hand side of the equation has an eigenvalue of unity; this occurs at the critical temperature $T_c$. Clearly, since Eqn.~(\ref{gap2}) is dimensionless, if there is a solution, it occurs at $\widetilde{T}$ of order unity, i.e. $T_c = J \widetilde{T}_c$ where $\widetilde{T}_c = O(1)$. The only remaining question is then whether a solution exists at any $\widetilde{T}_c > 0$. 

To show that such a solution indeed exists, we do not have to solve the equations explicitly. It suffices to show that in the limit $\widetilde{T} \rightarrow 0$, the largest eigenvalue of $M$ diverges; since the eigenvalues of $M$ all go to zero in the opposite limit $\widetilde{T}\rightarrow \infty$, this implies that the largest eigenvalue of $M$ has to cross unity at a finite value of $\widetilde{T}$.

In the regime $\widetilde{\omega},\widetilde{T}\ll 1$, we can replace $\widetilde{G}_i(\widetilde{\omega})$ by its form in the ``scaling regime'' of the SYK model: $\widetilde{G}_i(\widetilde{\omega})\sim i\tn{sgn}(\widetilde{\omega})/\sqrt{|\widetilde{\omega}|}$. Similarly, in this regime, $\widetilde{\Pi}(\widetilde{\Omega}) \sim -\log(1/\max(|\widetilde{\Omega}|,\widetilde{T}))$. Inserting these expressions in (\ref{gap2}) we obtain:
\begin{eqnarray}
\widetilde{\Delta}(\widetilde{\omega}) &=& \widetilde{T}\sum_{\widetilde{\Omega}} \frac{\widetilde{\Delta}~(\widetilde{\Omega})}{\sqrt{|\widetilde{\Omega}\widetilde{\omega}|}}\log\bigg[\frac{1}{\max(|\widetilde{\Omega}-\widetilde{\omega}|,\widetilde{T})}\bigg] \nonumber \\
&\equiv&
\sum_{\widetilde{\Omega}} \widetilde{M}(\widetilde{\omega},\widetilde{\Omega})\widetilde{\Delta}(\widetilde{\Omega}),
\label{eq:M}     
\end{eqnarray}
where we have defined $\widetilde{\Delta}(\widetilde{\omega}) = \Delta(\omega)/\sqrt{J|\widetilde{\omega}|}$, and the second line defines the symmetric matrix $\widetilde{M}$.

\begin{figure}[t]
\begin{center}
\includegraphics[scale=0.5]{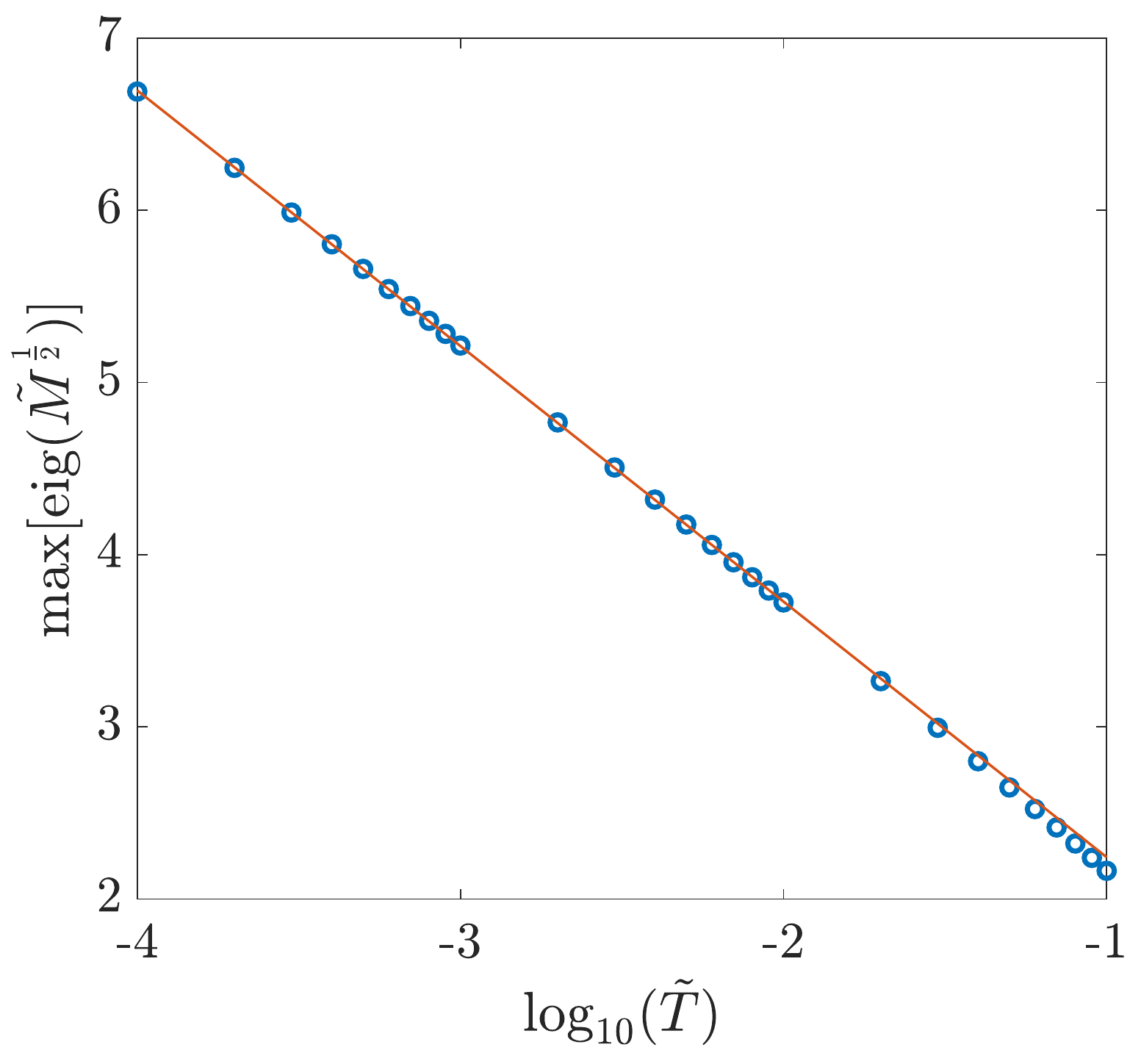}
\end{center}
\vspace{-8mm}
\caption{The maximum eigenvalue of $\widetilde{M}^{\frac{1}{2}}$ as a function of $\log_{10}(\widetilde{T})$ [see Eqs.~(\ref{eq:M},\ref{eq:eig})].}
\label{fig:eigs}
\end{figure}

We can now show that the largest eigenvalues of $\widetilde{M}$ diverge in the limit $\widetilde{T}\rightarrow 0$; to see this, one may use a trial solution $\widetilde{\Delta}_{\text{tr}}(\widetilde{\Omega}) = \Theta(1-|\widetilde{\Omega}|)/|\widetilde{\Omega}|^{1/2}$ and compute 
\begin{equation}
\frac{\sum_{\widetilde{\omega},\widetilde{\Omega}} \widetilde{\Delta}_{\text{tr}}(\widetilde{\omega})\widetilde{M}(\widetilde{\omega},\widetilde{\Omega})\widetilde{\Delta}_{\text{tr}}(\widetilde{\Omega}))}{\sum_{\widetilde{\omega}} |\widetilde{\Delta}_{\text{tr}}(\widetilde{\omega})|^2}\sim \log^2\left(\frac{1}{\widetilde{T}} \right).\label{eq:eig}    
\end{equation}
Fig.~\ref{fig:eigs} shows the scaling of the largest eigenvalue of $\widetilde{M}$ as a function of $\widetilde{T}$, confirming Eqn.~(\ref{eq:eig}). We conclude that Eqn.~(\ref{gap1}) indeed has a solution at a non-zero temperature $T_c \sim J$.

\begin{figure}[h]
\begin{center}
\includegraphics[scale=0.35]{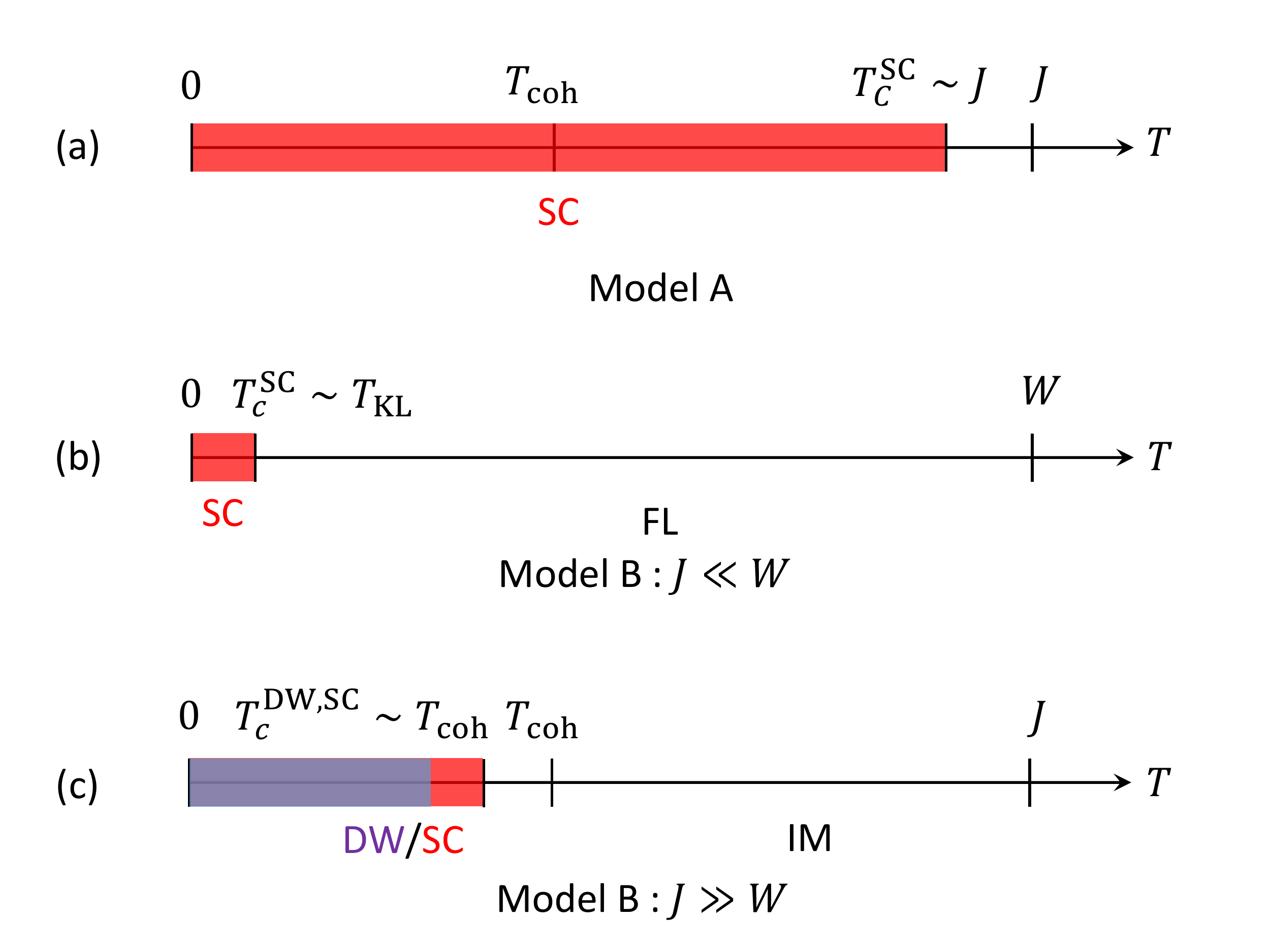}
\end{center}
\caption{Phase-diagrams for (a) Model$-\A$ ($J_{ijkl}=J_{ikjl}$) at small $U$. (b), (c) Model$-\B$ ($J_{ijkl}=-J_{ikjl}$) at weak ($J\ll W$) and strong ($J\gg W$) coupling, respectively. The superconducting/density-wave transition temperatures are denoted $T_c^{\tn{SC,DW}}$. The coherence scale, $T_{\tn{coh}}$, denotes the crossover scale between the incoherent metal (IM) and Fermi liquid (FL). At weak-coupling, $T^{\tn{SC}}_c$ is determined by the  Kohn-Luttinger (KL) scale.}
\label{pd}
\end{figure}

The resulting superconducting state has an on-site, spin-singlet `$s$-wave' pairing symmetry. However, in order to obtain a long-range ordered superconductor with a finite phase-stiffness, we need to include the effect of the inter-site single electron hopping terms, which can be treated perturbatively for $T\gg T_{\tn{coh}}$. Physically, it is clear that the system can be treated as local islands of superconductivity that are coupled to each other through an effective Josephson coupling ${\cal{J}}\sim Nt^2/J \gg J$. Thus, $T_c$ is indeed set by $J$, while the Josephson coupling (or, superfluid stiffness) is set by a much larger scale that is proportional to $N T_{\tn{coh}}$. In this case, there is no FL regime at low temperatures (see Fig.~\ref{pd}a). We should emphasize that the pairing instability in this case is unrelated to the usual `BCS-log', arising from the perfect nesting of states near the Fermi surface at $\pm\k$; instead it arises from the completely incoherent excitations in the locally critical non-Fermi liquid metal. Finally, we can address the fate of this instability for $U\neq0$. Clearly, in Eqn.~\ref{BS1a}, the pair-hopping term suppresses on-site pairing. When $U\gtrsim J$, the superconducting instability of this incoherent regime can be suppressed altogether.  

A generalization of the model to the one-band SYK lattice model with on-site $q-$fermion interactions (for $q>4$) shows similar physics, as discussed in Appendix~\ref{sykq}.

\section{Model$-\B$: Generalized Kohn-Luttinger mechanism}
\label{MB}

It is clear from the discussion in the previous section that in model-$\B$, as a result of the sign of $\overline{J_{ijkl} J_{ikjl}} = - J^2$, the corresponding Bethe-Salpeter equation in the incoherent regime (i.e. Eqn.~\ref{BS1b}) does {\it not} have a non-trivial solution at the level of a single site. It is then natural to ask if an instability can arise once we include the perturbative effects of the single-electron hopping, $t$.

It is well known that in conventional Fermi-liquid metals with purely repulsive interactions ($U_c>0$), an effective attraction can be generated in a non s-wave angular momentum channel at higher order in the interaction strength (i.e. $O(U_c^2)$ and higher). This traditionally goes under the name of `Kohn-Luttinger' (KL) effect \cite{KL} and typically leads to a small, $T_c\sim T_{\tn{KL}} \sim W\tn{exp}[-1/(\nu_0U_c)^2]$. At weak-coupling ($U_c\ll W$), the crucial ingredient responsible for the generation of an effective attraction relies on the momentum dependent structure of the particle-hole susceptibility, $\Pi(\k)$, for $|\k|\leq 2k_F$ that is generated self-consistently. Recall the peculiarity that in two-dimensions and for electrons with a parabolic dispersion $\ve_\k=\k^2/2m - \mu$, the susceptibility $\Pi(\k)$ is a constant for $|\k|< 2k_F$; this is no longer the case in higher dimensions and leads to a sign-changing (p-wave) gap function \cite{KL}. In two-dimensions, electrons with a parabolic dispersion exhibit a Kohn-Luttinger pairing instability at $O(U_c^3)$ \cite{AVC93}.

Within the incoherent metal regime of our model ($T\gg W^*$), the hopping leads to a non-singular, perturbative correction to the electron Green's function (Eqn.~\ref{limits}), which by itself can not lead to a diverging susceptibility. However, this does not preclude the possibility of an enhancement in the pairing susceptibility as a function of decreasing temperatures. In order to estimate the scale associated with this enhancement, we can revisit Eqn.~\ref{BS1b} and include the effects of a finite (but small) $t$. The lowest order contribution to the pairing vertex is at $O(t^2)$. Inserting Eq.~(\ref{limits}) for the Green's function at frequency $\omega \gg W^*$ into the Bethe-Salpeter equation (\ref{BS2a}) and performing a simple power-counting argument would suggest that the correction is given by,
\beq
\delta\chi_{\tn{pair}} \sim J^2\int_{T}^{J} d^2\omega~\frac{t^2}{J^3\omega^3}\sim \frac{t^2}{J}\bigg(\frac{1}{T} - \frac{1}{J}\bigg) \stackrel{J\rightarrow\infty}{\sim} \frac{T_{\tn{coh}}}{T},\nonumber\\
\eeq
where we have taken $J\rightarrow\infty$ while keeping $T_{\tn{coh}}\sim t^2/J$ finite. Clearly, when $T\gg T_{\tn{coh}}$, the correction is small. On the other hand, the above correction becomes $O(1)$ when $T$ approaches $T_{\tn{coh}}$ from above, which is the scale at which $t$ can no longer be treated perturbatively. Therefore, the above simple analysis already suggests that while the incoherent excitations in the NFL regime of model-$\B$ are unable to give rise to a pairing instability, it is likely that upon approaching the characteristic scale of $T_{\tn{coh}}$ from above, the system has an increased propensity towards developing superconductivity. In order to investigate this matter further we will now study the instabilities of model$-\B$ from the other asymptotic limit, namely low temperatures in the FL regime.  

\subsection{Strong coupling analysis}

Before proceeding further, let us revisit the saddle-point Eqn.~\ref{saddle} and the Bethe-Salpeter Eqn.~\ref{BS1b}, to cast them in a more transparent fashion. For the FL regime, which is of main interest in this section, the only relevant energy scale is the renormalized bandwidth, $W^*$. We thus carry out the following scaling transformations: $\widetilde\omega = \omega/W^*$,  $\widetilde{\ve}_\k = \ve_\k/W$ and $\widetilde{\Sigma}(\k,i\omega) = \Sigma(\k,i\omega)/W$. The physics is best elucidated in the strong-coupling limit with $J\rightarrow\infty$ and $T_{\tn{coh}}\sim W^* = W^2/J$ finite. We then obtain,
\beq
\widetilde{\Sigma}(\k,i\widetilde\omega) &=& -\int_{\k_1}\int_{\widetilde{\omega}_1} \widetilde{G}(\k_1,i\widetilde\omega_1)~\widetilde\Pi(\k+\k_1,i\widetilde\omega+i\widetilde\omega_1),\nonumber\\
\widetilde\Pi(\q,i\widetilde\Omega) &=& \int_{\k}\int_{\widetilde\omega} \widetilde{G}(\k,i\widetilde\omega)~\widetilde{G}(\k+\q,i\widetilde\omega+i\widetilde\Omega),
\label{scaledsaddle}
\eeq
where
\beq
\widetilde{G}(\k,i\widetilde\omega) = \frac{1}{-\widetilde\ve_\k - \widetilde\Sigma(\k,i\widetilde\omega)},
\label{scaledG}
\eeq
and $\widetilde{\Pi}(\widetilde{\Omega}) = \widetilde{T}\sum_{\widetilde{\omega}} \widetilde{G}_i(i\omega) \widetilde{G}_i(i\widetilde{\Omega}+i\widetilde{\omega})$. Notice that we have dropped the bare $`i\omega$' term in the Green's function, which is justified in the strong-coupling limit, where the entire frequency dependent renormalization arises from the singular frequency dependence of $\widetilde\Sigma$. We have thus been able to cast the original saddle-point equations in a purely dimensionless form. In this non-dimensional form, the crossover between the NFL and the FL regime occurs at $\widetilde\omega\sim 1$ (i.e. $\omega\sim W^*$). The Bethe-Salpeter Eqn.~\ref{BS1b} in terms of these scaled variables takes the form (we have set $U=0$),
\begin{widetext}
\beq
\widetilde\Delta_\l(\k,\widetilde\omega) =  \widetilde{T}\sum_{\widetilde\Omega}\int_\q\widetilde\Delta_i(\q,\widetilde\Omega)~\widetilde{G}_i(\q,i\widetilde\Omega)~\widetilde{G}_i(-\q,-i\widetilde\Omega)~\widetilde\Pi(\k-\q,i\widetilde\omega-i\widetilde\Omega),
\label{scaledBS}
\eeq
\end{widetext}
where we have also rescaled  $\widetilde{T}=T/W^*$, $\widetilde\Delta_\el(\k,\widetilde{\omega}) = \Delta_\el(\k,\widetilde\omega)/W$ and $\widetilde\Pi(\q,i\widetilde\Omega)$ is defined as before. Notice that for the linearized Bethe-Salpeter equation, there is no reason apriori to scale $\Delta_\el$ as above; for internal consistency, it is appropriate to scale $\Delta_\el$ and $\Sigma$ in the same fashion. Based on these rescalings, it is clear that if there is a solution to Eq.~(\ref{scaledBS}) in the strong coupling regime as defined above (with $J$ taken to infinity while keeping $W^*$ fixed), it occurs at $\widetilde{T}\sim O(1)$, i.e. $T_c= W^* \widetilde{T}_c \sim W^*$. We are now left with the task of investigating whether $\widetilde\Pi$ has the right momentum dependent structure to give any solution with $\widetilde{T}_c>0$.

To investigate whether a solution exists, we repeat the procedure introduced in Sec.~\ref{MA}: viewing Eq.~(\ref{scaledBS}) as a matrix equation, we ask whether the largest eigenvalue of the linear operator defined by the RHS has an eigenvalue that exceeds $1$ at a non-zero $\widetilde{T}$. Since all the eigenvalues becomes small when $\tilde{T}\gg 1$, it is sufficient to show that the largest eigenvalue diverges in the limit $\widetilde{T}\rightarrow 0$. To show this, note that in the low-temperature limit, the normal state becomes a Fermi liquid~\cite{DCsyk}, albeit with a strongly renormalized quasiparticle weight, $Z\sim W/J$ [see Eq.~(\ref{limits})]. This translates into $\widetilde{G}_i \approx \frac{1}{i\widetilde{\omega} - \widetilde{\varepsilon}_\k}$. Hence, in the limit of $\widetilde{\omega},\widetilde{\Omega} \ll 1$, Eq.~(\ref{scaledBS}) resembles the gap equation for a weakly interacting Fermi liquid. Then, so long as the density of states corresponding to the renormalized dispersion $\widetilde{\varepsilon}_\k$ is non-zero, we know that the magnitude of the largest eigenvalue of the RHS of (\ref{scaledBS}) diverges logarithmically in the limit $\widetilde{T}\rightarrow 0$, if the effective interaction given by $\widetilde{\Pi}(\q, \widetilde{\Omega}\rightarrow 0)$ has a positive (attractive) sign in a certain symmetry channel; this is just the usual Cooper logarithm. The only remaining question is regarding the sign of the interaction. For model-$\B$, if $\Delta$ does not depend on $\k$ (s-wave like order parameter), then all the eigenvalues are non-positive. However, solutions can arise in a non s-wave, anisotropic channel. 

In the low-frequency limit, Eq.~(\ref{scaledBS}) is identical to the Bethe-Salpeter equation that arises in the weak coupling treatment of the Hubbard model~\cite{Scalapino1986,Raghu10}. Hence, we know, e.g., that on a square lattice and when the density of electrons is close to half filling, there are non-trivial solutions, with the leading one having d-wave ($B_{1g}$) symmetry~\cite{Raghu10}. For our case, this implies that there is a non-trivial solution to Eq.~(\ref{scaledBS}). We stress, however, that the solution actually has $T_c \sim W^*$, and therefore cannot be regarded as an instability of the renormalized Fermi liquid; it appears at the same energy scale as the renormalized Fermi energy (see Fig.~\ref{pd}). Superconductivity in this model appears in the crossover regime between the incoherent metal and the Fermi liquid, and preempts the development of coherent quasi-particles.

\subsection{Explicit example: partially polarized Fermi sea}
\label{pfs}

Let us demonstrate the superconducting instability in model-$\B$ by considering an explicit tractable example. We modify the non-interacting part of the Hamiltonian in Eqn.~\ref{ham}. We consider in two-dimensions the presence of an external Zeeman field, $h$, that leads to a partial spin polarization. 
This leads to a modification of the dispersion for every orbital in a spin-dependent way ($s=\uparrow, \downarrow$): $\ve_{\k,is}\rightarrow \ve_{\k,is} + sh$. It is useful to consider the limit where we are near the bottom of the bands such that we can make a `parabolic' approximation (with the bare mass, $m\sim 1/ta^2$) for the dispersion and treat the problem as rotationally invariant,
\beq
\ve_{\k,is} = \frac{\k^2}{2m} + sh.
\eeq
For the non-interacting problem, the orbitals have a spin-dependent Fermi momentum, $k_{F,s}$, and Fermi velocity, $v_{F,s}=k_{F,s}/m$. The total density of states at the Fermi energy is given by, $\nu_{0,s}=\nu_0=m/2\pi$. The effect of a weak repulsive interaction on pairing was studied for a similar partially spin-polarized Fermi sea (with $N=1$ orbital) in Ref.~\cite{raghu}. Our focus in this paper is to study the effect of $H_J$ in Eqn.~\ref{hu} in the presence of spin-polarization and in the large$-N$ limit, at both weak and strong coupling.

It is worth pointing out that the strong-coupling limit for this particular case is slightly different.\footnote{Under the parabolic approximation, the kinetic energy is no longer bounded, i.e. the bandwidth is not finite.} We are interested in the limit: $J/\ve_F\gg1$ with $J\rightarrow\infty$ and $\ve_F\rightarrow\infty$, while keeping $k_F$ (i.e. the density) fixed; $\ve_F$ is the Fermi energy. For model$-\B$, the Schwinger-Dyson equations (Fig.~\ref{se}) in the presence of spin-polarization preserves the basic structure of the original setup. In particular, at low temperatures ($T\ll T_{\tn{coh}}$), the Green's function is still of the form given in Eqn.~\ref{scaledG}, where the coherence scale now is $T_{\tn{coh}}\sim \ve_F^2/J$, which we keep finite in the strong-coupling limit. We will primarily be interested in the physics at temperatures below $T_{\tn{coh}}$. The nature of the incoherent metallic state above $T_{\tn{coh}}$ in the present model with an unbounded bandwidth is an interesting question that we leave for the future. Below $T_{\tn{coh}}$, there is a crossover into an incipient FL regime, where the self-energy is momentum independent. Turning now to the possible instabilities, since the Fermi surfaces are spin-polarized in the FL regime, the leading superconducting instability (if any) will be in the spin-triplet channel. The spin-triplet, orbital-diagonal vertex is defined as,
\beq
\Delta_{is}(\r-\r') \equiv \langle  c_{\r is} c_{\r' is}\rangle.
\eeq

The Bethe-Salpeter equations for the two pairing vertices are given by (we suppress the orbital indices below),
\begin{widetext}
\beq
\Delta_\ua(\k,\omega) &=& -J^2 \sum_\Omega \int_\q \Delta_\ua(\q,\Omega)~ G_\ua(\q,i\Omega)~G_\ua(-\q,-i\Omega)~\Pi_\da(\k-\q,i\omega-i\Omega),\\
\Delta_\da(\k,\omega) &=& -J^2 \sum_\Omega \int_\q \Delta_\da(\q,\Omega)~ G_\da(\q,i\Omega)~G_\da(-\q,-i\Omega)~\Pi_\ua(\k-\q,i\omega-i\Omega),
\eeq
\end{widetext}
where the polarization bubble is defined as earlier,
\beq
\Pi_s(\k,i\omega) = T\sum_{\Omega}\int_\q G_s(\q,i\Omega)~G_s(\q+\k,i\omega+i\Omega).
\eeq

Let us first recall the results in the weak coupling limit, following the analysis of Ref.~\cite{raghu}. In this limit, we may ignore the frequency dependence of the gap functions~\cite{Raghu10}, and it is enough to study the pairing generated by the static part of the polarization function, which is given by
\beq
\Pi_s(\k) = \frac{Z\nu_0}{2}\bigg(1 - \frac{\tn{Re}\sqrt{k^2 - (2k_{F,s})^2}}{k} \bigg).
\eeq

Decomposing the triplet eigenfunctions as,
\beq
\Delta_s(\hat{k}) = \Delta(k_{F,s}) \cos(m\theta_{\hat{k}}), ~~m\in \tn{odd integers},
\eeq
leads to the following set of eigenvalue problems
\beq
\lambda_{m,\ua} &=& -Z\nu_0 J^2\int\frac{d\theta}{2\pi}~ \Pi_\da(2k_{F,\ua}|\sin(\theta/2)|)~\cos(m\theta),\nonumber\\ \\
\lambda_{m,\da} &=& -Z\nu_0 J^2\int\frac{d\theta}{2\pi}~ \Pi_\ua(2k_{F,\da}|\sin(\theta/2)|)~\cos(m\theta),\nonumber\\
\eeq
with $\theta$ the angle between $\k$ and $\q$. Just as in Ref.~\cite{raghu}, if we assume without any loss of generality that $k_{F\da}<k_{F\ua}$, $\lambda_{m,\da}=0$ for all $m$. The absence of an instability arises from the momentum independent structure of $\Pi_\ua(\hat{k}-\hat{q})$ for the appropriate momentum transfer on the smaller Fermi surface. On the other hand, the momentum transfer on the larger Fermi surface leads to an effective interaction, $\Pi_\da(\hat{k}-\hat{q})$, that can lead to a non-trivial solution. The corresponding eigenvalue is given by,
\beq
\lambda_{m,\ua}=\frac{Z^2\nu_0^2J^2}{\pi}\int_{\theta_c}^\pi~d\theta \frac{\sqrt{\sin^2(\theta/2) - \alpha^2}}{\sin(\theta/2)}~\cos(m\theta),
\eeq
where $0\leq\alpha(=k_{F,\da}/k_{F,\ua})\leq 1$ and $\theta_c=2\sin^{-1}\alpha$. In the $p-$wave channel, the above leads to,
\beq
\lambda_{1,\ua}(\alpha) = -Z^2\nu_0^2J^2\alpha(1-\alpha),
\eeq
which is non-zero as long as $\alpha\neq0,1$ (i.e. either fully polarized/ unpolarized). The transition temperature for superconductivity is then $T_c^{\tn{SC}}\sim  T_{\tn{coh}}~\tn{exp}(-1/|\lambda_{1,\ua}|)$. At weak-coupling, this is  
the celebrated Kohn-Luttinger result (with $T_{\tn{coh}}\rightarrow \ve_F$), except now the bare interaction at $O(J)$ is neither attractive nor repulsive and a net attraction is only generated at $O(J^2)$. 

We now turn to the regime of strong-coupling, which is still described by the same set of Schwinger-Dyson equation for the self-energy and Bethe-Salpeter equation for the pairing vertices at large$-N$. In particular, the quasiparticle residue in the FL regime is small, $Z\nu_0J\sim 1$, and correspondingly the pairing eigenvalue in the spin-triplet channel becomes $\lambda_{1,\ua}\sim O(1)$.  Thus, when we start from the low temperature description of the FL and focus on the states near the Fermi surface, we are led to the surprising result that the scale of the superconducting instability in the regime of strong-coupling is $T_c^{\tn{SC}}\sim T_{\tn{coh}}$.

Based on the above explicit example, as well as the structure of the Bethe-Salpeter equations, we can now make a few general observations. We have argued in the previous sections that if we start with the description of a low temperature Fermi liquid and focus on the excitations near the Fermi surface, then purely as a result of the strong renormalization of the coherent excitations by the factor of $Z\sim t/J$, {\footnote{For the model with a parabolic dispersion in the presence of spin-polarization, $t\rightarrow\ve_F$.}} the predicted $T_c^{\tn{SC}}\sim W^*\sim T_{\tn{coh}}$. However, at these scales that are comparable to the renormalized bandwidth, many of our underlying assumptions are not strictly applicable. In particular, the scattering rate is large, $\Sigma''(0,T)\sim W^*$, and it is not sufficient to focus only on the low-energy states near the Fermi surface. The Fermi surface itself is no longer sharply defined. Strictly speaking, the quasiparticle residue may no longer be treated as frequency independent all the way to $\omega\sim W^*$ and the contribution of the incoherent part of the spectral function must be included to study the onset of pairing. This is beyond the scope of this work. However, the appearance of $W^*$ as a special scale at which the pairing susceptibility exhibits non-trivial behavior, extrapolating both from the low-temperature Fermi liquid and the high-temperature incoherent metal, is suggestive of the fact that $T_c^{\tn{SC}}$ is indeed determined by $W^*$.

\section{Discussion and Outlook}
\label{disc}
 
In this work, we have studied a family of large-$N$ lattice models that display pairing instabilities beyond the conventional BCS framework. In particular, we would like to emphasize that the superconducting instabilities obtained for models$-\A$ and $\B$ in this work do not rely on the conventional `Cooper-log' arising from the nesting between time-reversed pairs of momenta on the Fermi surface.{\footnote{For model$-\B$, the conventional Cooper-log ensures, however, that there exists a pairing instability.}} In both cases, the superconducting state develops from a non-Fermi liquid state with no long lived quasiparticles; in model$-\A$, the parent normal state is a completely incoherent, locally quantum critical metal, whereas in model$-\B$, the normal state is at the crossover between the incoherent metal and the Fermi liquid phase, such that the coherent quasiparticles have not yet fully developed. Interestingly, in the latter case, the only scale in the Eliashberg equation (which is {\it exact} in the large$-N$ limit) is the renormalized bandwidth $W^*$. We therefore find that $T_c$ is of the order of $W^*$. This is somewhat reminiscent of the situation near a quantum critical point, where the only scale in the Eliashberg equation is the coupling constant~\cite{abanov2001coherent,Roussev2001}. However, we stress that in our model$-\B$, this situation arises without the need to tune to a quantum critical point. Similar physics occurs in model$-\A$ if a strong on-site repulsive ``pair hopping'' interaction $U$ is added (see Eq.~\ref{hp} above). 

The fact that the superconducting instability in our model does not rely on the existence of coherent quasiparticles near the Fermi surface, and hence does not require a degeneracy between opposite momenta, naturally leads to the question of whether the same interactions can induce other competing instabilities, for instance in the particle-hole channel. A careful consideration of the Bethe-Salpeter equations in the particle-hole channel immediately leads to the following conclusions: (i) The disorder contraction at $O(J^2)$ that appears in the particle-hole vertex (the particle-hole analogue of the pairing vertex in Fig.~\ref{pv}) is of the form $\overline{J_{ijkl}J_{ijkl}}$, and is thus insensitive to the precise nature of the permutation symmetries, as was the case for pairing. There is no difference between models$-\A$ and $\B$, as far as the instabilities in the particle-hole channel are concerned. (ii) The incoherent metal is stable against any instability in the particle-hole channel (i.e. there is no `on-site' instability). (iii) Finally, at weak coupling, there is no instability in the particle-hole channel; there is only a superconducting instability due to the usual Kohn-Luttinger mechanism. However, in the strong-coupling limit, {\it in principle} there could be an instability to density-wave order with $T_c^{\tn{DW}}\sim T_c^{\tn{SC}}\sim W^*$. All of our previous arguments for the pairing instability are applicable to also the particle-hole vertex, especially in the regime where the system can no longer be described as a weakly interacting Fermi liquid with a sharp Fermi surface as $T\rightarrow T_{\tn{coh}}$. However, as discussed above, while a pairing instability is guaranteed in this regime, whose origin can ultimately be traced to the usual `Cooper-log', this is not the case for the instability in the particle-hole channel. The precise nature of the density-wave instability,  vis-\`{a}-vis its (commensurate) wavevector, intra-unit cell form-factor (if a bond-density), and the numerical ratio of $T_c^{\tn{SC}}/T_c^{\tn{DW}} (\gtrsim1)$ is determined by underlying microscopic details of the starting Hamiltonian. We note that the above family of models is thus an interesting playground for exploring various {\it intertwined} orders at strong-coupling, where different orders all appear with the same basic energy scale $W^*$. We leave a detailed analysis of these questions, especially the possible instabilities in the particle-hole channel, to the future. 

A separate, but equally interesting question that has not been addressed in the present paper is the fate of superconducting instabilities in the two-band generalizations of the spinless version of the above model, where the bare bandwidth of one of the bands is much narrower than the other~\cite{DCsyk}. These two-band models realize critical Fermi surfaces with a `marginal' (or `non') Fermi liquid self-energies for the electrons with the larger bandwidth. It is natural to ask if the same interactions that destroy the long-lived quasiparticles near the critical Fermi surface can simultaneously induce pairing. For quantum critical metals, the precise answer to this question can depend on underlying details \cite{MM15}. We leave a detailed analysis of these and other related questions for the lattice-SYK model for the future.

\acknowledgements 
We thank A. Chubukov and T. Senthil for useful discussions. The research of DC at MIT was supported by a fellowship from the Gordon and Betty Moore
Foundation, under the EPiQS initiative, Grant GBMF-4303. EB was supported by the European Research Council (ERC) under grant HQMAT (grant no. 817799). DC acknowledges hospitality of the Weizmann Institute of Science and the Max-Planck Institute (PKS), where this work was completed.

{\it Note added:} While this manuscript was being finalized for submission, two manuscripts \cite{YW,JS} appeared on the arXiv that study pairing instabilites of a different variant of the $(0+1)-$dimensional SYK model.
  
\bibliographystyle{apsrev4-1_custom}

\bibliography{nfl}

\begin{thebibliography}{37}%
\makeatletter
\providecommand \@ifxundefined [1]{%
 \@ifx{#1\undefined}
}%
\providecommand \@ifnum [1]{%
 \ifnum #1\expandafter \@firstoftwo
 \else \expandafter \@secondoftwo
 \fi
}%
\providecommand \@ifx [1]{%
 \ifx #1\expandafter \@firstoftwo
 \else \expandafter \@secondoftwo
 \fi
}%
\providecommand \natexlab [1]{#1}%
\providecommand \enquote  [1]{``#1''}%
\providecommand \bibnamefont  [1]{#1}%
\providecommand \bibfnamefont [1]{#1}%
\providecommand \citenamefont [1]{#1}%
\providecommand \href@noop [0]{\@secondoftwo}%
\providecommand \href [0]{\begingroup \@sanitize@url \@href}%
\providecommand \@href[1]{\@@startlink{#1}\@@href}%
\providecommand \@@href[1]{\endgroup#1\@@endlink}%
\providecommand \@sanitize@url [0]{\catcode `\\12\catcode `\$12\catcode
  `\&12\catcode `\#12\catcode `\^12\catcode `\_12\catcode `\%12\relax}%
\providecommand \@@startlink[1]{}%
\providecommand \@@endlink[0]{}%
\providecommand \url  [0]{\begingroup\@sanitize@url \@url }%
\providecommand \@url [1]{\endgroup\@href {#1}{\urlprefix }}%
\providecommand \urlprefix  [0]{URL }%
\providecommand \Eprint [0]{\href }%
\providecommand \doibase [0]{http://dx.doi.org/}%
\providecommand \selectlanguage [0]{\@gobble}%
\providecommand \bibinfo  [0]{\@secondoftwo}%
\providecommand \bibfield  [0]{\@secondoftwo}%
\providecommand \translation [1]{[#1]}%
\providecommand \BibitemOpen [0]{}%
\providecommand \bibitemStop [0]{}%
\providecommand \bibitemNoStop [0]{.\EOS\space}%
\providecommand \EOS [0]{\spacefactor3000\relax}%
\providecommand \BibitemShut  [1]{\csname bibitem#1\endcsname}%
\let\auto@bib@innerbib\@empty
\bibitem [{\citenamefont {Keimer}\ \emph {et~al.}(2015)\citenamefont {Keimer},
  \citenamefont {Kivelson}, \citenamefont {Norman}, \citenamefont {Uchida},\
  and\ \citenamefont {Zaanen}}]{Keimer15}%
  \BibitemOpen
  \bibfield  {author} {\bibinfo {author} {\bibfnamefont {B.}~\bibnamefont
  {Keimer}}, \bibinfo {author} {\bibfnamefont {S.~A.}\ \bibnamefont
  {Kivelson}}, \bibinfo {author} {\bibfnamefont {M.~R.}\ \bibnamefont
  {Norman}}, \bibinfo {author} {\bibfnamefont {S.}~\bibnamefont {Uchida}}, \
  and\ \bibinfo {author} {\bibfnamefont {J.}~\bibnamefont {Zaanen}},\
  }\bibfield  {title} {\enquote {\bibinfo {title} {From quantum matter to
  high-temperature superconductivity in copper oxides},}\ }\href
  {http://dx.doi.org/10.1038/nature14165} {\bibfield  {journal} {\bibinfo
  {journal} {Nature}\ }\textbf {\bibinfo {volume} {518}},\ \bibinfo {pages}
  {179} (\bibinfo {year} {2015})}\BibitemShut {NoStop}%
\bibitem [{\citenamefont {Scalapino}(2012)}]{scalapino}%
  \BibitemOpen
  \bibfield  {author} {\bibinfo {author} {\bibfnamefont {D.~J.}\ \bibnamefont
  {Scalapino}},\ }\bibfield  {title} {\enquote {\bibinfo {title} {A common
  thread: The pairing interaction for unconventional superconductors},}\ }\href
  {\doibase 10.1103/RevModPhys.84.1383} {\bibfield  {journal} {\bibinfo
  {journal} {Rev. Mod. Phys.}\ }\textbf {\bibinfo {volume} {84}},\ \bibinfo
  {pages} {1383} (\bibinfo {year} {2012})}\BibitemShut {NoStop}%
\bibitem [{\citenamefont {Stewart}(2001)}]{Stewart}%
  \BibitemOpen
  \bibfield  {author} {\bibinfo {author} {\bibfnamefont {G.~R.}\ \bibnamefont
  {Stewart}},\ }\bibfield  {title} {\enquote {\bibinfo {title}
  {Non-fermi-liquid behavior in $d$- and $f$-electron metals},}\ }\href
  {\doibase 10.1103/RevModPhys.73.797} {\bibfield  {journal} {\bibinfo
  {journal} {Rev. Mod. Phys.}\ }\textbf {\bibinfo {volume} {73}},\ \bibinfo
  {pages} {797} (\bibinfo {year} {2001})}\BibitemShut {NoStop}%
\bibitem [{\citenamefont {Damascelli}\ \emph {et~al.}(2003)\citenamefont
  {Damascelli}, \citenamefont {Hussain},\ and\ \citenamefont {Shen}}]{zxs}%
  \BibitemOpen
  \bibfield  {author} {\bibinfo {author} {\bibfnamefont {A.}~\bibnamefont
  {Damascelli}}, \bibinfo {author} {\bibfnamefont {Z.}~\bibnamefont {Hussain}},
  \ and\ \bibinfo {author} {\bibfnamefont {Z.-X.}\ \bibnamefont {Shen}},\
  }\bibfield  {title} {\enquote {\bibinfo {title} {Angle-resolved photoemission
  studies of the cuprate superconductors},}\ }\href {\doibase
  10.1103/RevModPhys.75.473} {\bibfield  {journal} {\bibinfo  {journal} {Rev.
  Mod. Phys.}\ }\textbf {\bibinfo {volume} {75}},\ \bibinfo {pages} {473}
  (\bibinfo {year} {2003})}\BibitemShut {NoStop}%
\bibitem [{\citenamefont {Wang}\ \emph {et~al.}(2004)\citenamefont {Wang},
  \citenamefont {Yang}, \citenamefont {Sekharan}, \citenamefont {Ding},
  \citenamefont {Engelbrecht}, \citenamefont {Dai}, \citenamefont {Wang},
  \citenamefont {Kaminski}, \citenamefont {Valla}, \citenamefont {Kidd},
  \citenamefont {Fedorov},\ and\ \citenamefont {Johnson}}]{johnson}%
  \BibitemOpen
  \bibfield  {author} {\bibinfo {author} {\bibfnamefont {S.-C.}\ \bibnamefont
  {Wang}}, \bibinfo {author} {\bibfnamefont {H.-B.}\ \bibnamefont {Yang}},
  \bibinfo {author} {\bibfnamefont {A.~K.~P.}\ \bibnamefont {Sekharan}},
  \bibinfo {author} {\bibfnamefont {H.}~\bibnamefont {Ding}}, \bibinfo {author}
  {\bibfnamefont {J.~R.}\ \bibnamefont {Engelbrecht}}, \bibinfo {author}
  {\bibfnamefont {X.}~\bibnamefont {Dai}}, \bibinfo {author} {\bibfnamefont
  {Z.}~\bibnamefont {Wang}}, \bibinfo {author} {\bibfnamefont {A.}~\bibnamefont
  {Kaminski}}, \bibinfo {author} {\bibfnamefont {T.}~\bibnamefont {Valla}},
  \bibinfo {author} {\bibfnamefont {T.}~\bibnamefont {Kidd}}, \bibinfo {author}
  {\bibfnamefont {A.~V.}\ \bibnamefont {Fedorov}}, \ and\ \bibinfo {author}
  {\bibfnamefont {P.~D.}\ \bibnamefont {Johnson}},\ }\bibfield  {title}
  {\enquote {\bibinfo {title} {Quasiparticle line shape of
  ${\mathrm{sr}}_{2}{\mathrm{ruo}}_{4}$ and its relation to anisotropic
  transport},}\ }\href {\doibase 10.1103/PhysRevLett.92.137002} {\bibfield
  {journal} {\bibinfo  {journal} {Phys. Rev. Lett.}\ }\textbf {\bibinfo
  {volume} {92}},\ \bibinfo {pages} {137002} (\bibinfo {year}
  {2004})}\BibitemShut {NoStop}%
\bibitem [{\citenamefont {Takagi}\ \emph {et~al.}(1992)\citenamefont {Takagi},
  \citenamefont {Batlogg}, \citenamefont {Kao}, \citenamefont {Kwo},
  \citenamefont {Cava}, \citenamefont {Krajewski},\ and\ \citenamefont
  {Peck}}]{Takagi}%
  \BibitemOpen
  \bibfield  {author} {\bibinfo {author} {\bibfnamefont {H.}~\bibnamefont
  {Takagi}}, \bibinfo {author} {\bibfnamefont {B.}~\bibnamefont {Batlogg}},
  \bibinfo {author} {\bibfnamefont {H.~L.}\ \bibnamefont {Kao}}, \bibinfo
  {author} {\bibfnamefont {J.}~\bibnamefont {Kwo}}, \bibinfo {author}
  {\bibfnamefont {R.~J.}\ \bibnamefont {Cava}}, \bibinfo {author}
  {\bibfnamefont {J.~J.}\ \bibnamefont {Krajewski}}, \ and\ \bibinfo {author}
  {\bibfnamefont {W.~F.}\ \bibnamefont {Peck}},\ }\bibfield  {title} {\enquote
  {\bibinfo {title} {Systematic evolution of temperature-dependent resistivity
  in
  ${\mathrm{la}}_{2\mathrm{\ensuremath{-}}\mathit{x}}$${\mathrm{sr}}_{\mathit{x}}$${\mathrm{cuo}}_{4}$},}\
  }\href {\doibase 10.1103/PhysRevLett.69.2975} {\bibfield  {journal} {\bibinfo
   {journal} {Phys. Rev. Lett.}\ }\textbf {\bibinfo {volume} {69}},\ \bibinfo
  {pages} {2975} (\bibinfo {year} {1992})}\BibitemShut {NoStop}%
\bibitem [{\citenamefont {Marel}\ \emph {et~al.}(2003)\citenamefont {Marel},
  \citenamefont {Molegraaf}, \citenamefont {Zaanen}, \citenamefont {Nussinov},
  \citenamefont {Carbone}, \citenamefont {Damascelli}, \citenamefont {Eisaki},
  \citenamefont {Greven}, \citenamefont {Kes},\ and\ \citenamefont
  {Li}}]{Marel}%
  \BibitemOpen
  \bibfield  {author} {\bibinfo {author} {\bibfnamefont {D.~v.~d.}\
  \bibnamefont {Marel}}, \bibinfo {author} {\bibfnamefont {H.~J.~A.}\
  \bibnamefont {Molegraaf}}, \bibinfo {author} {\bibfnamefont {J.}~\bibnamefont
  {Zaanen}}, \bibinfo {author} {\bibfnamefont {Z.}~\bibnamefont {Nussinov}},
  \bibinfo {author} {\bibfnamefont {F.}~\bibnamefont {Carbone}}, \bibinfo
  {author} {\bibfnamefont {A.}~\bibnamefont {Damascelli}}, \bibinfo {author}
  {\bibfnamefont {H.}~\bibnamefont {Eisaki}}, \bibinfo {author} {\bibfnamefont
  {M.}~\bibnamefont {Greven}}, \bibinfo {author} {\bibfnamefont {P.~H.}\
  \bibnamefont {Kes}}, \ and\ \bibinfo {author} {\bibfnamefont
  {M.}~\bibnamefont {Li}},\ }\bibfield  {title} {\enquote {\bibinfo {title}
  {Quantum critical behaviour in a high-tc superconductor},}\ }\href
  {https://doi.org/10.1038/nature01978} {\bibfield  {journal} {\bibinfo
  {journal} {Nature}\ }\textbf {\bibinfo {volume} {425}},\ \bibinfo {pages}
  {271} (\bibinfo {year} {2003})}\BibitemShut {NoStop}%
\bibitem [{\citenamefont {Li}\ \emph {et~al.}(2004)\citenamefont {Li},
  \citenamefont {Taillefer}, \citenamefont {Hawthorn}, \citenamefont {Tanatar},
  \citenamefont {Paglione}, \citenamefont {Sutherland}, \citenamefont {Hill},
  \citenamefont {Wang},\ and\ \citenamefont {Chen}}]{Taillefer1}%
  \BibitemOpen
  \bibfield  {author} {\bibinfo {author} {\bibfnamefont {S.~Y.}\ \bibnamefont
  {Li}}, \bibinfo {author} {\bibfnamefont {L.}~\bibnamefont {Taillefer}},
  \bibinfo {author} {\bibfnamefont {D.~G.}\ \bibnamefont {Hawthorn}}, \bibinfo
  {author} {\bibfnamefont {M.~A.}\ \bibnamefont {Tanatar}}, \bibinfo {author}
  {\bibfnamefont {J.}~\bibnamefont {Paglione}}, \bibinfo {author}
  {\bibfnamefont {M.}~\bibnamefont {Sutherland}}, \bibinfo {author}
  {\bibfnamefont {R.~W.}\ \bibnamefont {Hill}}, \bibinfo {author}
  {\bibfnamefont {C.~H.}\ \bibnamefont {Wang}}, \ and\ \bibinfo {author}
  {\bibfnamefont {X.~H.}\ \bibnamefont {Chen}},\ }\bibfield  {title} {\enquote
  {\bibinfo {title} {Giant electron-electron scattering in the fermi-liquid
  state of
  ${\mathrm{n}\mathrm{a}}_{0.7}\mathrm{C}\mathrm{o}{\mathrm{o}}_{2}$},}\ }\href
  {\doibase 10.1103/PhysRevLett.93.056401} {\bibfield  {journal} {\bibinfo
  {journal} {Phys. Rev. Lett.}\ }\textbf {\bibinfo {volume} {93}},\ \bibinfo
  {pages} {056401} (\bibinfo {year} {2004})}\BibitemShut {NoStop}%
\bibitem [{\citenamefont {Hussey}\ \emph {et~al.}(1998)\citenamefont {Hussey},
  \citenamefont {Mackenzie}, \citenamefont {Cooper}, \citenamefont {Maeno},
  \citenamefont {Nishizaki},\ and\ \citenamefont {Fujita}}]{hussey}%
  \BibitemOpen
  \bibfield  {author} {\bibinfo {author} {\bibfnamefont {N.~E.}\ \bibnamefont
  {Hussey}}, \bibinfo {author} {\bibfnamefont {A.~P.}\ \bibnamefont
  {Mackenzie}}, \bibinfo {author} {\bibfnamefont {J.~R.}\ \bibnamefont
  {Cooper}}, \bibinfo {author} {\bibfnamefont {Y.}~\bibnamefont {Maeno}},
  \bibinfo {author} {\bibfnamefont {S.}~\bibnamefont {Nishizaki}}, \ and\
  \bibinfo {author} {\bibfnamefont {T.}~\bibnamefont {Fujita}},\ }\bibfield
  {title} {\enquote {\bibinfo {title} {Normal-state magnetoresistance of
  ${\mathrm{sr}}_{2}{\mathrm{ruo}}_{4}$},}\ }\href {\doibase
  10.1103/PhysRevB.57.5505} {\bibfield  {journal} {\bibinfo  {journal} {Phys.
  Rev. B}\ }\textbf {\bibinfo {volume} {57}},\ \bibinfo {pages} {5505}
  (\bibinfo {year} {1998})}\BibitemShut {NoStop}%
\bibitem [{\citenamefont {Hussey}\ \emph {et~al.}(2004)\citenamefont {Hussey},
  \citenamefont {Takenaka},\ and\ \citenamefont {Takagi}}]{Hussey04}%
  \BibitemOpen
  \bibfield  {author} {\bibinfo {author} {\bibfnamefont {N.~E.}\ \bibnamefont
  {Hussey}}, \bibinfo {author} {\bibfnamefont {K.}~\bibnamefont {Takenaka}}, \
  and\ \bibinfo {author} {\bibfnamefont {H.}~\bibnamefont {Takagi}},\
  }\bibfield  {title} {\enquote {\bibinfo {title} {Universality of the mott
  ioffe regel limit in metals},}\ }\href {\doibase
  10.1080/14786430410001716944} {\bibfield  {journal} {\bibinfo  {journal}
  {Philosophical Magazine}\ }\textbf {\bibinfo {volume} {84}},\ \bibinfo
  {pages} {2847} (\bibinfo {year} {2004})}\BibitemShut {NoStop}%
\bibitem [{\citenamefont {Emery}\ and\ \citenamefont {Kivelson}(1995)}]{SAK95}%
  \BibitemOpen
  \bibfield  {author} {\bibinfo {author} {\bibfnamefont {V.~J.}\ \bibnamefont
  {Emery}}\ and\ \bibinfo {author} {\bibfnamefont {S.~A.}\ \bibnamefont
  {Kivelson}},\ }\bibfield  {title} {\enquote {\bibinfo {title}
  {Superconductivity in bad metals},}\ }\href {\doibase
  10.1103/PhysRevLett.74.3253} {\bibfield  {journal} {\bibinfo  {journal}
  {Phys. Rev. Lett.}\ }\textbf {\bibinfo {volume} {74}},\ \bibinfo {pages}
  {3253} (\bibinfo {year} {1995})}\BibitemShut {NoStop}%
\bibitem [{\citenamefont {Sachdev}\ and\ \citenamefont {Ye}(1993)}]{SY}%
  \BibitemOpen
  \bibfield  {author} {\bibinfo {author} {\bibfnamefont {S.}~\bibnamefont
  {Sachdev}}\ and\ \bibinfo {author} {\bibfnamefont {J.}~\bibnamefont {Ye}},\
  }\bibfield  {title} {\enquote {\bibinfo {title} {Gapless spin-fluid ground
  state in a random quantum heisenberg magnet},}\ }\href {\doibase
  10.1103/PhysRevLett.70.3339} {\bibfield  {journal} {\bibinfo  {journal}
  {Phys. Rev. Lett.}\ }\textbf {\bibinfo {volume} {70}},\ \bibinfo {pages}
  {3339} (\bibinfo {year} {1993})}\BibitemShut {NoStop}%
\bibitem [{\citenamefont {Kitaev}()}]{kitaev_talk}%
  \BibitemOpen
  \bibfield  {author} {\bibinfo {author} {\bibfnamefont {A.}~\bibnamefont
  {Kitaev}},\ }\bibfield  {title} {\enquote {\bibinfo {title} {A simple model
  of quantum holography, talk given at kitp program: entanglement in
  strongly-correlated quantum matter},}\ }\href@noop {} {\bibinfo  {journal}
  {USA April 2015}\ }\BibitemShut {NoStop}%
\bibitem [{\citenamefont {Maldacena}\ \emph {et~al.}(2016)\citenamefont
  {Maldacena}, \citenamefont {Shenker},\ and\ \citenamefont
  {Stanford}}]{Maldacena2016}%
  \BibitemOpen
\bibfield  {journal} {  }\bibfield  {author} {\bibinfo {author} {\bibfnamefont
  {J.}~\bibnamefont {Maldacena}}, \bibinfo {author} {\bibfnamefont {S.~H.}\
  \bibnamefont {Shenker}}, \ and\ \bibinfo {author} {\bibfnamefont
  {D.}~\bibnamefont {Stanford}},\ }\bibfield  {title} {\enquote {\bibinfo
  {title} {A bound on chaos},}\ }\href {\doibase 10.1007/JHEP08(2016)106}
  {\bibfield  {journal} {\bibinfo  {journal} {Journal of High Energy Physics}\
  }\textbf {\bibinfo {volume} {2016}},\ \bibinfo {pages} {106} (\bibinfo {year}
  {2016})}\BibitemShut {NoStop}%
\bibitem [{\citenamefont {{Kitaev}}\ and\ \citenamefont
  {{Suh}}(2017)}]{kitaevsuh}%
  \BibitemOpen
  \bibfield  {author} {\bibinfo {author} {\bibfnamefont {A.}~\bibnamefont
  {{Kitaev}}}\ and\ \bibinfo {author} {\bibfnamefont {S.~J.}\ \bibnamefont
  {{Suh}}},\ }\bibfield  {title} {\enquote {\bibinfo {title} {{The soft mode in
  the Sachdev-Ye-Kitaev model and its gravity dual}},}\ }\href@noop {}
  {\bibfield  {journal} {\bibinfo  {journal} {ArXiv e-prints}\ } (\bibinfo
  {year} {2017})},\ \Eprint {http://arxiv.org/abs/1711.08467} {arXiv:1711.08467
  [hep-th]} \BibitemShut {NoStop}%
\bibitem [{\citenamefont {Parcollet}\ and\ \citenamefont
  {Georges}(1999)}]{Parcollet1}%
  \BibitemOpen
  \bibfield  {author} {\bibinfo {author} {\bibfnamefont {O.}~\bibnamefont
  {Parcollet}}\ and\ \bibinfo {author} {\bibfnamefont {A.}~\bibnamefont
  {Georges}},\ }\bibfield  {title} {\enquote {\bibinfo {title}
  {Non-fermi-liquid regime of a doped mott insulator},}\ }\href {\doibase
  10.1103/PhysRevB.59.5341} {\bibfield  {journal} {\bibinfo  {journal} {Phys.
  Rev. B}\ }\textbf {\bibinfo {volume} {59}},\ \bibinfo {pages} {5341}
  (\bibinfo {year} {1999})}\BibitemShut {NoStop}%
\bibitem [{\citenamefont {Georges}\ \emph {et~al.}(2001)\citenamefont
  {Georges}, \citenamefont {Parcollet},\ and\ \citenamefont
  {Sachdev}}]{Parcollet2}%
  \BibitemOpen
  \bibfield  {author} {\bibinfo {author} {\bibfnamefont {A.}~\bibnamefont
  {Georges}}, \bibinfo {author} {\bibfnamefont {O.}~\bibnamefont {Parcollet}},
  \ and\ \bibinfo {author} {\bibfnamefont {S.}~\bibnamefont {Sachdev}},\
  }\bibfield  {title} {\enquote {\bibinfo {title} {Quantum fluctuations of a
  nearly critical heisenberg spin glass},}\ }\href {\doibase
  10.1103/PhysRevB.63.134406} {\bibfield  {journal} {\bibinfo  {journal} {Phys.
  Rev. B}\ }\textbf {\bibinfo {volume} {63}},\ \bibinfo {pages} {134406}
  (\bibinfo {year} {2001})}\BibitemShut {NoStop}%
\bibitem [{\citenamefont {Gu}\ \emph {et~al.}(2017)\citenamefont {Gu},
  \citenamefont {Qi},\ and\ \citenamefont {Stanford}}]{Gu17}%
  \BibitemOpen
  \bibfield  {author} {\bibinfo {author} {\bibfnamefont {Y.}~\bibnamefont
  {Gu}}, \bibinfo {author} {\bibfnamefont {X.-L.}\ \bibnamefont {Qi}}, \ and\
  \bibinfo {author} {\bibfnamefont {D.}~\bibnamefont {Stanford}},\ }\bibfield
  {title} {\enquote {\bibinfo {title} {Local criticality, diffusion and chaos
  in generalized sachdev-ye-kitaev models},}\ }\href {\doibase
  10.1007/JHEP05(2017)125} {\bibfield  {journal} {\bibinfo  {journal} {Journal
  of High Energy Physics}\ }\textbf {\bibinfo {volume} {2017}},\ \bibinfo
  {pages} {125} (\bibinfo {year} {2017})}\BibitemShut {NoStop}%
\bibitem [{\citenamefont {Davison}\ \emph {et~al.}(2017)\citenamefont
  {Davison}, \citenamefont {Fu}, \citenamefont {Georges}, \citenamefont {Gu},
  \citenamefont {Jensen},\ and\ \citenamefont {Sachdev}}]{SS17}%
  \BibitemOpen
  \bibfield  {author} {\bibinfo {author} {\bibfnamefont {R.~A.}\ \bibnamefont
  {Davison}}, \bibinfo {author} {\bibfnamefont {W.}~\bibnamefont {Fu}},
  \bibinfo {author} {\bibfnamefont {A.}~\bibnamefont {Georges}}, \bibinfo
  {author} {\bibfnamefont {Y.}~\bibnamefont {Gu}}, \bibinfo {author}
  {\bibfnamefont {K.}~\bibnamefont {Jensen}}, \ and\ \bibinfo {author}
  {\bibfnamefont {S.}~\bibnamefont {Sachdev}},\ }\bibfield  {title} {\enquote
  {\bibinfo {title} {Thermoelectric transport in disordered metals without
  quasiparticles: The sachdev-ye-kitaev models and holography},}\ }\href
  {\doibase 10.1103/PhysRevB.95.155131} {\bibfield  {journal} {\bibinfo
  {journal} {Phys. Rev. B}\ }\textbf {\bibinfo {volume} {95}},\ \bibinfo
  {pages} {155131} (\bibinfo {year} {2017})}\BibitemShut {NoStop}%
\bibitem [{\citenamefont {Song}\ \emph {et~al.}(2017)\citenamefont {Song},
  \citenamefont {Jian},\ and\ \citenamefont {Balents}}]{Balents}%
  \BibitemOpen
  \bibfield  {author} {\bibinfo {author} {\bibfnamefont {X.-Y.}\ \bibnamefont
  {Song}}, \bibinfo {author} {\bibfnamefont {C.-M.}\ \bibnamefont {Jian}}, \
  and\ \bibinfo {author} {\bibfnamefont {L.}~\bibnamefont {Balents}},\
  }\bibfield  {title} {\enquote {\bibinfo {title} {Strongly correlated metal
  built from sachdev-ye-kitaev models},}\ }\href {\doibase
  10.1103/PhysRevLett.119.216601} {\bibfield  {journal} {\bibinfo  {journal}
  {Phys. Rev. Lett.}\ }\textbf {\bibinfo {volume} {119}},\ \bibinfo {pages}
  {216601} (\bibinfo {year} {2017})}\BibitemShut {NoStop}%
\bibitem [{\citenamefont {Zhang}(2017)}]{Zhang17}%
  \BibitemOpen
  \bibfield  {author} {\bibinfo {author} {\bibfnamefont {P.}~\bibnamefont
  {Zhang}},\ }\bibfield  {title} {\enquote {\bibinfo {title} {Dispersive
  sachdev-ye-kitaev model: Band structure and quantum chaos},}\ }\href
  {\doibase 10.1103/PhysRevB.96.205138} {\bibfield  {journal} {\bibinfo
  {journal} {Phys. Rev. B}\ }\textbf {\bibinfo {volume} {96}},\ \bibinfo
  {pages} {205138} (\bibinfo {year} {2017})}\BibitemShut {NoStop}%
\bibitem [{\citenamefont {Patel}\ \emph
  {et~al.}(2018{\natexlab{a}})\citenamefont {Patel}, \citenamefont {McGreevy},
  \citenamefont {Arovas},\ and\ \citenamefont {Sachdev}}]{SSmagneto}%
  \BibitemOpen
  \bibfield  {author} {\bibinfo {author} {\bibfnamefont {A.~A.}\ \bibnamefont
  {Patel}}, \bibinfo {author} {\bibfnamefont {J.}~\bibnamefont {McGreevy}},
  \bibinfo {author} {\bibfnamefont {D.~P.}\ \bibnamefont {Arovas}}, \ and\
  \bibinfo {author} {\bibfnamefont {S.}~\bibnamefont {Sachdev}},\ }\bibfield
  {title} {\enquote {\bibinfo {title} {Magnetotransport in a model of a
  disordered strange metal},}\ }\href {\doibase 10.1103/PhysRevX.8.021049}
  {\bibfield  {journal} {\bibinfo  {journal} {Phys. Rev. X}\ }\textbf {\bibinfo
  {volume} {8}},\ \bibinfo {pages} {021049} (\bibinfo {year}
  {2018}{\natexlab{a}})}\BibitemShut {NoStop}%
\bibitem [{\citenamefont {{Khveshchenko}}(2017)}]{DVK17}%
  \BibitemOpen
  \bibfield  {author} {\bibinfo {author} {\bibfnamefont {D.~V.}\ \bibnamefont
  {{Khveshchenko}}},\ }\bibfield  {title} {\enquote {\bibinfo {title}
  {{Thickening and sickening the SYK model}},}\ }\href@noop {} {\bibfield
  {journal} {\bibinfo  {journal} {ArXiv e-prints}\ } (\bibinfo {year}
  {2017})},\ \Eprint {http://arxiv.org/abs/1705.03956} {arXiv:1705.03956
  [cond-mat.str-el]} \BibitemShut {NoStop}%
\bibitem [{\citenamefont {Jian}\ \emph {et~al.}(2018)\citenamefont {Jian},
  \citenamefont {Xian},\ and\ \citenamefont {Yao}}]{hongyao}%
  \BibitemOpen
  \bibfield  {author} {\bibinfo {author} {\bibfnamefont {S.-K.}\ \bibnamefont
  {Jian}}, \bibinfo {author} {\bibfnamefont {Z.-Y.}\ \bibnamefont {Xian}}, \
  and\ \bibinfo {author} {\bibfnamefont {H.}~\bibnamefont {Yao}},\ }\bibfield
  {title} {\enquote {\bibinfo {title} {Quantum criticality and duality in the
  $\text{Sachdev-Ye-Kitaev}/{\mathrm{ads}}_{2}$ chain},}\ }\href {\doibase
  10.1103/PhysRevB.97.205141} {\bibfield  {journal} {\bibinfo  {journal} {Phys.
  Rev. B}\ }\textbf {\bibinfo {volume} {97}},\ \bibinfo {pages} {205141}
  (\bibinfo {year} {2018})}\BibitemShut {NoStop}%
\bibitem [{\citenamefont {Chowdhury}\ \emph {et~al.}(2018)\citenamefont
  {Chowdhury}, \citenamefont {Werman}, \citenamefont {Berg},\ and\
  \citenamefont {Senthil}}]{DCsyk}%
  \BibitemOpen
  \bibfield  {author} {\bibinfo {author} {\bibfnamefont {D.}~\bibnamefont
  {Chowdhury}}, \bibinfo {author} {\bibfnamefont {Y.}~\bibnamefont {Werman}},
  \bibinfo {author} {\bibfnamefont {E.}~\bibnamefont {Berg}}, \ and\ \bibinfo
  {author} {\bibfnamefont {T.}~\bibnamefont {Senthil}},\ }\bibfield  {title}
  {\enquote {\bibinfo {title} {Translationally invariant non-fermi-liquid
  metals with critical fermi surfaces: Solvable models},}\ }\href {\doibase
  10.1103/PhysRevX.8.031024} {\bibfield  {journal} {\bibinfo  {journal} {Phys.
  Rev. X}\ }\textbf {\bibinfo {volume} {8}},\ \bibinfo {pages} {031024}
  (\bibinfo {year} {2018})}\BibitemShut {NoStop}%
\bibitem [{\citenamefont {Patel}\ \emph
  {et~al.}(2018{\natexlab{b}})\citenamefont {Patel}, \citenamefont {Lawler},\
  and\ \citenamefont {Kim}}]{Patel}%
  \BibitemOpen
  \bibfield  {author} {\bibinfo {author} {\bibfnamefont {A.~A.}\ \bibnamefont
  {Patel}}, \bibinfo {author} {\bibfnamefont {M.~J.}\ \bibnamefont {Lawler}}, \
  and\ \bibinfo {author} {\bibfnamefont {E.-A.}\ \bibnamefont {Kim}},\
  }\bibfield  {title} {\enquote {\bibinfo {title} {Coherent superconductivity
  with a large gap ratio from incoherent metals},}\ }\href {\doibase
  10.1103/PhysRevLett.121.187001} {\bibfield  {journal} {\bibinfo  {journal}
  {Phys. Rev. Lett.}\ }\textbf {\bibinfo {volume} {121}},\ \bibinfo {pages}
  {187001} (\bibinfo {year} {2018}{\natexlab{b}})}\BibitemShut {NoStop}%
\bibitem [{\citenamefont {Kohn}\ and\ \citenamefont {Luttinger}(1965)}]{KL}%
  \BibitemOpen
  \bibfield  {author} {\bibinfo {author} {\bibfnamefont {W.}~\bibnamefont
  {Kohn}}\ and\ \bibinfo {author} {\bibfnamefont {J.~M.}\ \bibnamefont
  {Luttinger}},\ }\bibfield  {title} {\enquote {\bibinfo {title} {New mechanism
  for superconductivity},}\ }\href {\doibase 10.1103/PhysRevLett.15.524}
  {\bibfield  {journal} {\bibinfo  {journal} {Phys. Rev. Lett.}\ }\textbf
  {\bibinfo {volume} {15}},\ \bibinfo {pages} {524} (\bibinfo {year}
  {1965})}\BibitemShut {NoStop}%
\bibitem [{\citenamefont {Chubukov}(1993)}]{AVC93}%
  \BibitemOpen
  \bibfield  {author} {\bibinfo {author} {\bibfnamefont {A.~V.}\ \bibnamefont
  {Chubukov}},\ }\bibfield  {title} {\enquote {\bibinfo {title} {Kohn-luttinger
  effect and the instability of a two-dimensional repulsive fermi liquid at
  t=0},}\ }\href {\doibase 10.1103/PhysRevB.48.1097} {\bibfield  {journal}
  {\bibinfo  {journal} {Phys. Rev. B}\ }\textbf {\bibinfo {volume} {48}},\
  \bibinfo {pages} {1097} (\bibinfo {year} {1993})}\BibitemShut {NoStop}%
\bibitem [{\citenamefont {Scalapino}\ \emph {et~al.}(1986)\citenamefont
  {Scalapino}, \citenamefont {Loh},\ and\ \citenamefont
  {Hirsch}}]{Scalapino1986}%
  \BibitemOpen
  \bibfield  {author} {\bibinfo {author} {\bibfnamefont {D.~J.}\ \bibnamefont
  {Scalapino}}, \bibinfo {author} {\bibfnamefont {E.}~\bibnamefont {Loh}}, \
  and\ \bibinfo {author} {\bibfnamefont {J.~E.}\ \bibnamefont {Hirsch}},\
  }\bibfield  {title} {\enquote {\bibinfo {title} {$d$-wave pairing near a
  spin-density-wave instability},}\ }\href {\doibase 10.1103/PhysRevB.34.8190}
  {\bibfield  {journal} {\bibinfo  {journal} {Phys. Rev. B}\ }\textbf {\bibinfo
  {volume} {34}},\ \bibinfo {pages} {8190} (\bibinfo {year}
  {1986})}\BibitemShut {NoStop}%
\bibitem [{\citenamefont {Raghu}\ \emph {et~al.}(2010)\citenamefont {Raghu},
  \citenamefont {Kivelson},\ and\ \citenamefont {Scalapino}}]{Raghu10}%
  \BibitemOpen
  \bibfield  {author} {\bibinfo {author} {\bibfnamefont {S.}~\bibnamefont
  {Raghu}}, \bibinfo {author} {\bibfnamefont {S.~A.}\ \bibnamefont {Kivelson}},
  \ and\ \bibinfo {author} {\bibfnamefont {D.~J.}\ \bibnamefont {Scalapino}},\
  }\bibfield  {title} {\enquote {\bibinfo {title} {Superconductivity in the
  repulsive hubbard model: An asymptotically exact weak-coupling solution},}\
  }\href {\doibase 10.1103/PhysRevB.81.224505} {\bibfield  {journal} {\bibinfo
  {journal} {Phys. Rev. B}\ }\textbf {\bibinfo {volume} {81}},\ \bibinfo
  {pages} {224505} (\bibinfo {year} {2010})}\BibitemShut {NoStop}%
\bibitem [{\citenamefont {Raghu}\ and\ \citenamefont {Kivelson}(2011)}]{raghu}%
  \BibitemOpen
  \bibfield  {author} {\bibinfo {author} {\bibfnamefont {S.}~\bibnamefont
  {Raghu}}\ and\ \bibinfo {author} {\bibfnamefont {S.~A.}\ \bibnamefont
  {Kivelson}},\ }\bibfield  {title} {\enquote {\bibinfo {title}
  {Superconductivity from repulsive interactions in the two-dimensional
  electron gas},}\ }\href {\doibase 10.1103/PhysRevB.83.094518} {\bibfield
  {journal} {\bibinfo  {journal} {Phys. Rev. B}\ }\textbf {\bibinfo {volume}
  {83}},\ \bibinfo {pages} {094518} (\bibinfo {year} {2011})}\BibitemShut
  {NoStop}%
\bibitem [{\citenamefont {Abanov}\ \emph {et~al.}(2001)\citenamefont {Abanov},
  \citenamefont {Chubukov},\ and\ \citenamefont
  {Finkel'stein}}]{abanov2001coherent}%
  \BibitemOpen
  \bibfield  {author} {\bibinfo {author} {\bibfnamefont {A.}~\bibnamefont
  {Abanov}}, \bibinfo {author} {\bibfnamefont {A.~V.}\ \bibnamefont
  {Chubukov}}, \ and\ \bibinfo {author} {\bibfnamefont {A.}~\bibnamefont
  {Finkel'stein}},\ }\bibfield  {title} {\enquote {\bibinfo {title} {Coherent
  vs. incoherent pairing in 2d systems near magnetic instability},}\ }\href
  {https://iopscience.iop.org/article/10.1209/epl/i2001-00266-0/meta}
  {\bibfield  {journal} {\bibinfo  {journal} {EPL (Europhysics Letters)}\
  }\textbf {\bibinfo {volume} {54}},\ \bibinfo {pages} {488} (\bibinfo {year}
  {2001})}\BibitemShut {NoStop}%
\bibitem [{\citenamefont {Roussev}\ and\ \citenamefont
  {Millis}(2001)}]{Roussev2001}%
  \BibitemOpen
  \bibfield  {author} {\bibinfo {author} {\bibfnamefont {R.}~\bibnamefont
  {Roussev}}\ and\ \bibinfo {author} {\bibfnamefont {A.~J.}\ \bibnamefont
  {Millis}},\ }\bibfield  {title} {\enquote {\bibinfo {title} {Quantum critical
  effects on transition temperature of magnetically mediated p-wave
  superconductivity},}\ }\href {\doibase 10.1103/PhysRevB.63.140504} {\bibfield
   {journal} {\bibinfo  {journal} {Phys. Rev. B}\ }\textbf {\bibinfo {volume}
  {63}},\ \bibinfo {pages} {140504} (\bibinfo {year} {2001})}\BibitemShut
  {NoStop}%
\bibitem [{\citenamefont {Metlitski}\ \emph {et~al.}(2015)\citenamefont
  {Metlitski}, \citenamefont {Mross}, \citenamefont {Sachdev},\ and\
  \citenamefont {Senthil}}]{MM15}%
  \BibitemOpen
  \bibfield  {author} {\bibinfo {author} {\bibfnamefont {M.~A.}\ \bibnamefont
  {Metlitski}}, \bibinfo {author} {\bibfnamefont {D.~F.}\ \bibnamefont
  {Mross}}, \bibinfo {author} {\bibfnamefont {S.}~\bibnamefont {Sachdev}}, \
  and\ \bibinfo {author} {\bibfnamefont {T.}~\bibnamefont {Senthil}},\
  }\bibfield  {title} {\enquote {\bibinfo {title} {Cooper pairing in non-fermi
  liquids},}\ }\href {\doibase 10.1103/PhysRevB.91.115111} {\bibfield
  {journal} {\bibinfo  {journal} {Phys. Rev. B}\ }\textbf {\bibinfo {volume}
  {91}},\ \bibinfo {pages} {115111} (\bibinfo {year} {2015})}\BibitemShut
  {NoStop}%
\bibitem [{\citenamefont {{Wang}}(2019)}]{YW}%
  \BibitemOpen
  \bibfield  {author} {\bibinfo {author} {\bibfnamefont {Y.}~\bibnamefont
  {{Wang}}},\ }\bibfield  {title} {\enquote {\bibinfo {title} {{A Solvable
  Random Model With Quantum-critical Points for Non-Fermi-liquid Pairing}},}\
  }\href@noop {} {\bibfield  {journal} {\bibinfo  {journal} {arXiv e-prints}\
  ,\ \bibinfo {eid} {arXiv:1904.07240}} (\bibinfo {year} {2019})},\ \Eprint
  {http://arxiv.org/abs/1904.07240} {arXiv:1904.07240 [cond-mat.str-el]}
  \BibitemShut {NoStop}%
\bibitem [{\citenamefont {{Esterlis}}\ and\ \citenamefont
  {{Schmalian}}(2019)}]{JS}%
  \BibitemOpen
  \bibfield  {author} {\bibinfo {author} {\bibfnamefont {I.}~\bibnamefont
  {{Esterlis}}}\ and\ \bibinfo {author} {\bibfnamefont {J.}~\bibnamefont
  {{Schmalian}}},\ }\bibfield  {title} {\enquote {\bibinfo {title} {{Cooper
  pairing of incoherent electrons: an electron-phonon version of the
  Sachdev-Ye-Kitaev model}},}\ }\href@noop {} {\bibfield  {journal} {\bibinfo
  {journal} {arXiv e-prints}\ ,\ \bibinfo {eid} {arXiv:1906.04747}} (\bibinfo
  {year} {2019})},\ \Eprint {http://arxiv.org/abs/1906.04747} {arXiv:1906.04747
  [cond-mat.str-el]} \BibitemShut {NoStop}%
\bibitem [{\citenamefont {Gross}\ and\ \citenamefont
  {Rosenhaus}(2017)}]{Gross17}%
  \BibitemOpen
  \bibfield  {author} {\bibinfo {author} {\bibfnamefont {D.~J.}\ \bibnamefont
  {Gross}}\ and\ \bibinfo {author} {\bibfnamefont {V.}~\bibnamefont
  {Rosenhaus}},\ }\bibfield  {title} {\enquote {\bibinfo {title} {A
  generalization of sachdev-ye-kitaev},}\ }\href {\doibase
  10.1007/JHEP02(2017)093} {\bibfield  {journal} {\bibinfo  {journal} {Journal
  of High Energy Physics}\ }\textbf {\bibinfo {volume} {2017}},\ \bibinfo
  {pages} {93} (\bibinfo {year} {2017})}\BibitemShut {NoStop}%
\end{thebibliography}%

\begin{widetext}
\appendix

\section{Generalization to translationally invariant SYK$_q$ models}
\label{sykq}
In this appendix, we consider a generalization of the model originally introduced in Ref.~\cite{Gross17} and extended to a translationally invariant lattice in Ref.~\cite{DCsyk},
 \beq
H_c = \sum_{\r,\r'} \sum_{\el,s_1} (-t_{\r,\r'} { - \mu \delta_{\r\r'} }) c_{\r \el s_1}^\dagger c_{\r'\el s_1} + \frac{\left(q/2\right) !}{N^{\frac{q-1}{2}}}\sum_{\{i_\ell\}}\sum_{\{s_i\}} J_{i_1i_2...i_q} \bigg[c^\dagger_{\r,i_1 s_1}c^\dagger_{\r,i_2 s_2}...c^\dagger_{\r,i_{q/2}s_{q/2}}c_{\r,i_{q/2+1}s_{q/2}}...c_{\r,i_{q-1}s_2}c_{\r,i_q s_1} \bigg].\nonumber\\
\label{hc}
\eeq
As before we take $J_{i_1i_2...i_q}$ and the hopping $t$ to be translationally invariant, with $\overline{J_{i_1i_2...i_q}}= 0$, and $\overline{(J_{i_1i_2...i_q})^2}=J^2$. In analogy with model$-\A$ considered above, we impose additional correlations among the matrix elements of the form: $J_{i_1i_2...i_{q-1}i_q} = J_{i_1i_{q-1}...i_2i_q}$. Imposing this additional structure does not change the saddle-point solution; for general $q$, the scaling dimension of the fermion in the absence of the hopping term is $\Delta(q) = 1/q$. Following the earlier discussion, it is straightforward to see that the coherence scale $T_{\tn{coh}} = t\left(t/J\right)^{\frac{2}{q-2}}$, above which the metal exhibits a locally critical, incoherent regime. The gap equations (ignoring the momentum dependence) take the familiar form,
\beq
\Delta_\l(\omega) = -J^2T\sum_\Omega \Delta_i(\Omega)~ G_i(i\Omega)~G_i(-i\Omega)~\chi(i\omega-i\Omega),
\eeq
where $G_i(i\Omega)\sim i\tn{sgn}(\Omega)/(J^{2\Delta(q)} |\Omega|^{1-2\Delta(q)})$, and,
\beq
\chi(\tau)\sim [G(\tau)]^{(q-2)/2}~[G(-\tau)]^{(q-2)/2}.
\eeq
Assembling all of these constraints leads to,
\beq
\Delta_\l(\omega) = T\sum_\Omega \Delta_i(\Omega) \frac{1}{|\Omega|^{2-4\Delta(q)}} \frac{1}{|\omega-\Omega|^{4\Delta(q)-1}},
\eeq
where it is interesting to note that the coupling, $J$, has dropped out. It is then easy to see that just like in the $q=4$ case, the transition temperature, $T_c\sim J$, the only relevant scale in the problem. 

\end{widetext}
\end{document}